\documentclass[11pt]{article}
\usepackage[a4paper, margin=2.5cm]{geometry}

\usepackage[round]{natbib}
\usepackage{authblk}

\usepackage[english]{babel}
\usepackage{stmaryrd}
\usepackage[normalem]{ulem}
\usepackage{bm}

\usepackage[colorlinks=true,urlcolor=blue,citecolor=blue,linkcolor=blue,bookmarks=true]{hyperref}

\usepackage{multirow}
\usepackage{arydshln}
\usepackage{adjustbox}
\usepackage{amssymb,amsmath,amsthm,amscd}
\usepackage{mathrsfs}
\usepackage{frcursive}
\usepackage{xcolor}
\usepackage{textcomp}
\usepackage{dsfont}
\usepackage{lscape}
\usepackage{graphicx}
\usepackage{color}
\usepackage{frcursive}
\usepackage{paralist}

\usepackage[T1]{fontenc}
\usepackage[utf8]{inputenc}

\usepackage{float}
\usepackage{caption}
\usepackage{subcaption}
\usepackage{enumitem}

\theoremstyle{plain}
\newtheorem{theorem}{Theorem}
\newtheorem{lemma}{Lemma}

\newcommand{\dsum}{\displaystyle\sum}

\theoremstyle{remark}
\newtheorem{remark}{Remark}

\theoremstyle{definition}
\newtheorem{definition}{Definition}

\title{A functional spatial autoregressive model using signatures}
\author[,1]{Camille Frévent\thanks{Corresponding author: \texttt{camille.frevent@univ-lille.fr}}}
\affil[1]{Univ. Lille, CHU Lille, ULR 2694 - METRICS: Évaluation des technologies de santé et des pratiques médicales, F-59000 Lille, France.}
\date{}

\begin{document}

\maketitle   

\noindent \rule{\linewidth}{0.4pt} \\
We propose a new approach to the autoregressive spatial functional model, based on the notion of signature, which represents a function as an infinite series of its iterated integrals. It presents the advantage of being applicable to a wide range of processes.  \\
After having provided theoretical guarantees to the proposed model, we have shown in a simulation study and on a real data set that this new approach presents competitive performances compared to the traditional model. 
\rule{\linewidth}{0.4pt}

\section{Introduction}\label{sec:level1}

Thanks to progress in sensing and data storage capacities, data are increasingly being measured continuously over time. This led to the emergence and popularization of functional data analysis (FDA) by \cite{ramsaylivre}, and then to the adaptation of classical statistical methods to the functional framework \citep{reg_chiou,jacques2014model,delaigle2019clustering}.  In domains in which data naturally involve a spatial component (demography, environmental science, and agricultural science \citep{Hung_16}), the emergence
of functional data has led to the introduction of spatial functional data as well as new
methods for clustering \citep{romano2010clustering,vandewalle2022clustering}, kriging \citep{giraldo2011ordinary,ignaccolo2014kriging} or regression \citep{attouch2011robust,bernardi2017penalized}. See \cite{martinez2020recent} and \cite{mateu2021geostatistical} for recent reviews of methods used in spatial FDA. \\

We are interested here in spatial regression models for lattice data. A well-known and widely used model is the spatial autoregressive (SAR) model proposed by \cite{Cliff} which models the relationship between a real-valued response variable and real covariates by considering endogenous interactions. In the context of spatial functional data with $N$ spatial units, the objective is the modelling of the relationship between a real-valued random variable $Y$ and a functional covariate $\{X(t), t \in \mathcal{T}\}$ observed in the $N$ spatial locations. \\
In this context, \cite{huang2018spatial}, \cite{pineda2019functional} and \cite{ahmed2022quasi} assumed that $X$ belongs to $\mathcal{L}^{2}(\mathcal{T})$, the space of square-integrable functions on $\mathcal{T}$, and proposed to consider the following functional spatial autoregressive (FSAR) model:
$$
Y_i = \rho^* \sum_{j=1}^{N} v_{ij,N}Y_j + \int_{\mathcal{T}}X_i(t)\theta^*(t) \ \text{d}t + \varepsilon_i,\qquad i=1,\ldots,N,\, \; N=1,2,\ldots
$$ 
where the spatial dependency structure between the $N$ spatial units is described by an $N\times N$ non-stochastic spatial weights matrix $V_{N}=(v_{ij,N})_{1 \le i,j \le N}$ that depends on $N$, the autoregressive parameter $\rho^*$ is in a compact space $\mathcal{R}$ and $\theta^* \in \mathcal{L}^2(\mathcal{T})$. \\

In recent years, signatures have been widely used in many domains such as character recognition \citep{yang2015chinese, liu2017ps}, finance \citep{gyurko2013extracting, arribas2018derivatives}, recurrent neural networks \citep{lai2017online} and medicine \citep{kormilitzin2016application, perez2018signature, morrill2020utilization}. Initially defined by \cite{chen1957integration, chen1977iterated} for smooth paths and rediscovered
in the context of rough path theory \citep{lyons1998differential, friz2010multidimensional}, the signatures have been recently investigated by \cite{fermanian2022functional} in the context of a non-spatial linear regression model with functional covariates. They present the advantages of being applicable to a wide range of processes that are not necessarily square-integrable processes, and to better capture the differences between some kind of curves \citep{o1985curve,fermanian2022functional}. \\ This motivated us to explore here a new approach, based on the notion of signatures, to the traditional FSAR model. Section \ref{sec:2} presents the notion of signatures as well as the proposed signatures-based spatial autoregressive model, its estimation procedure and theoretical guaranties. Section \ref{sec:simu} describes a simulation study to compare our approach which the one of \cite{ahmed2022quasi}. Our method is then applied to a real data set in Section \ref{sec:appli}. Lastly, the results are discussed in Section \ref{sec:conclu}.

\section{The signatures-based spatial autoregressive model}\label{sec:2}

\subsection{Concept of signatures}

\noindent Let $\mathcal{T}$ be a compact interval and
		
$$
\begin{array}{cccl}
  X: & \mathcal{T} & \to & E       \\
  &   t & \to & \left(X_t^{(1)},...,X_t^{(p)} \right)   \\
\end{array}
$$
 be a $p-$dimensional path. Let the set of functions of finite $m-$variation on $\mathcal{T}$ denoted $$BV_m(\mathcal{T},E)=\{X:\mathcal{T} \to E,\;\| X \|_{TV,m}< \infty\},$$ 
 with 
$$\| X \|_{TV,m}=	\left(\sup_{I}\sum_{(t_0,...t_d)\in I} \| X_{t_i}-X_{t_{i-1}}\|^m\right)^{1/m},$$ where the supremum is over all finite partitions $I$ of $\mathcal{T}$ and $\|.\|$ is the Euclidean norm on $E$.  Let $\mathcal{C}_m(\mathcal{T},E)$ be the set of continuous path $X:\mathcal{T} \to E$ of finite $m-$variation.\\
		
\begin{definition}(Signature of a path; \cite{levin2013learning})\\
Let $X\in \mathcal{C}_m(\mathcal{T},E)$. The signature $Sig(X)$ of $X$ over the interval $\mathcal{T}$  is 
$$Sig(X)=(1,\mathbf{X}^1,\dots,\mathbf{X}^d,\dots)$$
where the following integral makes sense
$$
\mathbf{X}^d= \underset{\substack{
  t_1<\dots<t_d \\
t_1,...,t_d\in \mathcal{T}\\
}}{\int \dots \int} \ \text{d}X_{t_1}\otimes \ldots \otimes \text{d}X_{t_d} \in E^{\otimes d}.	$$
The truncated signature of order $D$ is $Sig^D(X)=(1,\mathbf{X}^1,\dots,\mathbf{X}^D)$, for every integer $D\ge 1$.
\end{definition}

In the following, we assume that $E=\mathbb{R}^p$, $p \ge 2$, $X\in \mathcal{C}_1(\mathcal{T},\mathbb{R}^p)$ and $\mathcal{T}=[0,1]$.
It should be noted that the assumption $X\in \mathcal{C}_1(\mathcal{T},\mathbb{R}^p)$ is less restrictive than assuming $X\in \mathcal{L}^{2}(\mathcal{T},\mathbb{R}^p)$. 

By using the convention $(\mathbb{R}^p)^{\otimes 0}=\mathbb{R}$, we define $T((\mathbb{R}^p))$ the tensor algebra space by
$$
T((\mathbb{R}^p)):=\left\{(a_0,a_1,\dots,a_d,\dots), \ \forall d\ge 0, a_d\in (\mathbb{R}^p)^{\otimes d} \right\}
$$ and $T^d((\mathbb{R}^p))$ the $d^{\text{th}}$ truncated tensor algebra space
$$
T^d((\mathbb{R}^p)):=\bigoplus_{i=0}^d (\mathbb{R}^p)^{\otimes i}.
$$
Then the signature $Sig(X)$ is valued in $T((\mathbb{R}^p))$ while the truncated version of order $D$, $Sig^D(X)$ is an element of $T^D((\mathbb{R}^p))$.\\

\begin{definition}(Signature coefficient of a path; \cite{levin2013learning}, Remark 2.5)\\
Let $A^*$ be the set of multi-indexes with entries in $\{1,\dots,p\}$. For $J\in A^*$ with length $|J|=d$, $J$ may be written $J=(i_1,\dots,i_d)$, with $i_j\in \{ 1, \dots, p \}$, $\forall j\in \{ 1, \dots, d \}$.\\
Let $(e_i)_{i=1}^p$ be the canonical orthonormal basis of $\mathbb{R}^p$. For any positive integer $d$, the space $(\mathbb{R}^p)^{\otimes d}$ is isomorphic to the free vector space generated by all the words of length $d$ in $A^*$
and $(e_{i_1}\otimes \dots \otimes e_{i_d})_{(i_1,\dots,i_d)\in \{ 1, \dots, p\}^d}$ form a basis of $(\mathbb{R}^p)^{\otimes d}$. Then the signature of $X$ can be rewritten as 
\begin{equation}
\label{signaturecoef}
Sig(X)=1+\sum_{d=1}^\infty \sum_{(i_1, \dots, i_d)} \mathcal{S}_{(i_1, \dots, i_d)}(X) e_{i_1}\otimes \dots \otimes e_{i_d}\in T((\mathbb{R}^p))
\end{equation}
where $$ \mathcal{S}_{(i_1,\dots,i_d)}(X)= \underset{\substack{
    t_1<\dots<t_d     \\
 t_1,\dots,t_d\in \mathcal{T} }}{\int \dots \int} \ \text{d}X_{t_1}^{(i_1)} \dots  \text{d}X_{t_d}^{(i_d)} \in \mathbb{R}. $$
\end{definition}

In the next sections we adopt the more conventional notation for functional data $$
\begin{array}{cccl}
  X: & \mathcal{T} & \to & \mathbb{R}^p       \\
  &   t & \to & \left(X^{(1)}(t),...,X_t^{(p)}(t)\right)   \\
\end{array}
.
$$

\subsection{Model}
Let us consider the following signatures-based spatial autoregressive model: 
\begin{equation}\label{modelsignature}
Y_i=\rho^* \sum_{j=1}^{N} v_{ij,N}Y_j + \alpha^* + \langle \theta^*,Sig(X_i) \rangle + \varepsilon_i
\end{equation}

where the autoregressive parameter $\rho^*$ is in a compact space $\mathcal{R}$ and the parameter $\theta^*$ is assumed to be written as an infinite series of tensors: 
$$ \theta^* = 1 + \dsum_{d=1}^{\infty} \dsum_{(i_1, \dots, i_d)} \beta_{(i_1, \dots, i_d)}^* e_{i_1}\otimes \dots \otimes e_{i_d}. $$

$V_N=(v_{ij,N})_{1\le i,j \le N}$ is a spatial weight matrix that it is common but not necessary to \textit{row normalize} in practice. This allows $0 \le v_{ij,N} \le 1$ and $-1 \le \rho^* \le 1$. In this way, the spatially-lagged variables are equal to a weighted average of the neighboring values. \\

The disturbances $\left\{\varepsilon_{i}, \; i=1,\ldots, N,\; N=1,2,\ldots \right\}$ are assumed to be independent and identically distributed random variables such that $\mathbb{E}(\varepsilon_{i})=0$ and $\mathbb{E}(\varepsilon_{i}^2)=\sigma^{2*}$. They are also independent of $\{X_i(t),t\in \mathcal{T}, \; i=1,\ldots,N,\; N=1,2\ldots\}$.\\

Then, one can rewrite (\ref{modelsignature}) as 
\begin{equation} \label{modelsignaturedevelop}
Y_i=\rho^*\sum_{j=1}^{N} v_{ij,N}Y_j + \alpha^* + 1 + \dsum_{d=1}^{\infty} \dsum_{(i_1, \dots, i_d)} \beta_{(i_1, \dots, i_d)}^* \mathcal{S}_{(i_1, \dots, i_d)}(X_i) + \varepsilon_i.
\end{equation}

However, this model cannot be maximized without addressing the difficulty produced by the infinite dimension of the signatures $Sig(X_i)$ (and thus the infinite number of coefficients $\beta_{(i_1, \dots, i_d)}^*$). The next two sections propose two estimation methods that overcome this challenge.

\subsubsection{Penalized signatures-based spatial regression (PenSSAR)}
Due to the infinite dimension of the signatures, $\rho^*$, $\alpha^*$, the coefficients $\beta_{(i_1, \dots, i_d)}^*$ and $\sigma^{2*}$ cannot be directly estimated. To solve this problem, we propose in this section a first strategy of estimation based on truncated signatures and a penalized (ridge) regression similarly to \cite{fermanian2022functional}. 

\paragraph{Estimation} \mbox{} \\

We consider here a slightly modified version of Model (\ref{modelsignaturedevelop}). We assume that the signature coefficients $\mathcal{S}_{(i_1, \dots, i_d)}(X_i)$ are involved in the model only up to a certain unknown truncation order $D^* \in \mathbb{N}$. The model thus becomes

\begin{equation}\label{modelsignaturetrunc}
Y_i=\rho^*\sum_{j=1}^{N} v_{ij,N}Y_j + \alpha^* + 1 + \dsum_{d=1}^{D^*} \dsum_{(i_1, \dots, i_d)} \beta_{(i_1, \dots, i_d)}^* \mathcal{S}_{(i_1, \dots, i_d)}(X_i) + \varepsilon_i.
\end{equation}

Let the signature coefficients vector of $X$, $\mathcal{S}(X)$, be the sequence of all signature coefficients:
$$\mathcal{S}(X)=(1,  \mathcal{S}_{(1)}(X),\dots,\mathcal{S}_{(p)}(X), \mathcal{S}_{(1,1)}(X),\mathcal{S}_{(1,2)}(X),\dots,\mathcal{S}_{(i_1,\dots,i_d)}(X),\dots) $$
and the truncated signature coefficients vector at order $D$ of $X$, $\mathcal{S}^D(X)$, be defined as
$$\mathcal{S}^D(X)=(1,  \mathcal{S}_{(1)}(X),\mathcal{S}_{(2)}(X),\dots,\mathcal{S}_{\underbrace{\scriptstyle (p,\dots,p)}_{D \text{ terms}}}(X)).$$

Then, by noting $s_p(D)=\sum_{d=0}^D p^d=\frac{p^{D+1}-1}{p-1}$ the dimension of the truncated signature coefficients vector at order $D$, (\ref{modelsignaturetrunc}) can be rewritten as
\begin{equation}\label{modelsignaturetrunc1}
S_N \mathbf{Y}_N = (\alpha^*+1) \mathbf{1}_N + \xi_{N}B^* + \bm{\varepsilon}_N,
\end{equation}

where $S_N=(I_N-\rho^* V_N)$, $\mathbf{Y}_N$ and $\bm{\varepsilon}_N$ are two $N\times 1$ vectors of elements $Y_i$ and $\varepsilon_i,\ i=1,\ldots,N$ respectively, $I_N$ denotes the $N\times N$ identity matrix and $\mathbf{1}_N$ denotes the $N \times 1$ vector composed only of 1. $B^*=(\beta_1^*,\beta_2^*, \ldots,\beta^*_{\underbrace{\scriptstyle (p,\dots,p)}_{D^* \text{ terms}}})^\top \in \mathbb{R}^{s_p(D^*)-1}$ and $\xi_{N}$ is an $N\times (s_p(D^*)-1)$ matrix whose $i^{\text{th}}$ line corresponds to $\mathcal{S}^{D^*}(X_i)$ without the constant 1. \\
Similarly, since $D^*$ is unknown, we define 
$B_{D}=(\beta_1,\beta_2, \ldots,\beta_{\underbrace{\scriptstyle (p,\dots,p)}_{D \text{ terms}}})^\top \in \mathbb{R}^{s_p(D)-1}$, and $\xi_{N,D}$ the $N\times (s_p(D)-1)$ matrix whose $i^{\text{th}}$ line corresponds to $\mathcal{S}^{D}(X_i)$ without the constant 1. Thus, $\xi_{N} = \xi_{N,D^*}$. 

Then, for a truncation order $D$, the associated conditional quasi log-likelihood function is
\begin{align}
\ell_N(\sigma^2,\rho,\alpha+1, B_{D})=&-\frac{N}{2}\ln(\sigma^2)-\frac{N}{2}\ln(2\pi)+  \ln\vert S_N(\rho)\vert \nonumber\\ & -\frac{1}{2\sigma^2}\left[S_N(\rho)\mathbf{Y}_N- (\alpha+1) \mathbf{1}_N - \xi_{N,D}B_{D}\right]^\top \left[S_N(\rho)\mathbf{Y}_N- (\alpha+1) \mathbf{1}_N - \xi_{N,D}B_{D}\right]
\label{TLK}
\end{align}

where $S_N(\rho)=I_N-\rho V_N$.

However, despite the use of truncated signatures, Model (\ref{modelsignaturetrunc1}) for a truncation order $D$ remains difficult to estimate from (\ref{TLK}) due to the number $s_p(D)-1$ components of $\xi_{N,D}$ which grows very quickly with $p$ and $D$. \\

Penalized regression allows the estimation of regression models with a large number of potentially correlated covariates. In particular, \cite{fermanian2022functional} proposed to estimate a functional (non-spatial) linear regression model with truncated signatures using a Ridge regularization. Here we propose to adapt the penalized regression algorithm proposed by \cite{liu2018penalized} in the context of our new model using a Ridge regularization. \\

More precisely, we propose the following iterative algorithm for a truncation order $D$:
\begin{enumerate}[label= \bfseries Step \arabic*,leftmargin=1.9cm]
\item Estimate the initial value for the estimation of the parameters $\alpha$, $B_D$ and $\sigma^{2}$: $\alpha^{(0)}$, $B_D^{(0)}$ and $\sigma^{2(0)}$ using a non-spatial Ridge regression. We note $\lambda_N$ the associated regularization parameter.
\item \label{step2} Update $\sigma^{2(m+1)} = \underset{\sigma^2>0}{\arg \max} \ \ell_N\left(\sigma^2, \rho^{(m)}, \alpha^{(m)}+1, B_D^{(m)} \right)$
\item Update $\rho^{(m+1)} = \underset{\rho \in \mathcal{R}}{\arg \max} \ \ell_N\left(\sigma^{2(m+1)}, \rho, \alpha^{(m)}+1, B_D^{(m)} \right)$
\item \label{step4} Update $\left(\alpha^{(m+1)}, B_D^{(m+1)^\top}\right) = \underset{(\alpha, B_D^\top) \in \mathbb{R}^{s_p(D)}}{\arg \max} \ \ell_N\left(\sigma^{2(m+1)}, \rho^{(m+1)}, \alpha+1, B_D \right) - N \lambda_N ||B_D||_2^2 $ \\
\item Repeat \ref{step2} to \ref{step4} until convergence \\
\end{enumerate}

This estimation procedure can be subject to several remarks.
\begin{remark}
In practice, $D^*$ corresponding to the truncation order $D$ of the ``true model'' (Model (\ref{modelsignaturetrunc1})) is unknown. Thus, the idea we propose is to apply the previous algorithm to a list of $D$ ranging from 1 to a maximum truncation order $D^{\text{max}} \in \mathbb{N}^*$, and then to select $D^*$ as the $D$ giving the smallest mean squared error by cross-validation.
\end{remark}
\begin{remark}
As this algorithm for estimating the spatial regression model (\ref{modelsignaturetrunc1}) is iterative, the computation time can be quite long. For this reason, it is rather challenging to estimate the optimal regularization parameter $\lambda_N$ using, for example, a grid and selecting the best one by cross-validation, as this would considerably increase computation time. Instead, we proposed to determine this parameter in the initialization step, using a non-spatial ridge regression, since the computation time for this approach is much more reasonable. Although the parameter obtained is therefore not the optimal one in the spatial case, we expect it not to be too far from it. 
\end{remark}
\begin{remark}
Moreover, it should be observed that by noting $\chi_{N,D}=(\mathbf{1}_N, \xi_{N,D})$ and $\gamma_{D}^{(m)} = (\alpha^{(m)}+1, B_{D}^{(m)\top})^\top$, 
the iterative estimations of $\sigma^{2*}$ and $(\alpha^*,B_D^{*\top})$ have explicit formulas at \ref{step2} and \ref{step4}:
$$ \sigma^{2(m+1)} = \dfrac{1}{N} \left[\left(I_N-\rho^{(m)} V_N \right) \mathbf{Y}_N - \chi_{N,D} \gamma_D^{(m)} \right]^\top\left[\left(I_N-\rho^{(m)} V_N \right) \mathbf{Y}_N - \chi_{N,D} \gamma_D^{(m)} \right] $$ and
$$ \gamma_D^{(m+1)} = \left[ \dfrac{1}{\sigma^{2(m+1)}} \chi_{N,D}^\top \chi_{N,D} + 2 N \Lambda_{N,D} \right]^{-1} \left[\dfrac{1}{\sigma^{2(m+1)}} \chi_{N,D}^\top \left(I_N - \rho^{(m+1)} V_N\right) \mathbf{Y}_N \right]$$
where $\Lambda_{N,D}=\begin{pmatrix} 0 & 0 & 0 & \dots & 0 \\ 0 & \lambda_N & 0 & \dots & 0 \\
0 & 0 & \ddots & \ddots & \vdots \\
\vdots & \ddots & \ddots & \ddots & 0 \\
0 & \dots & 0 & 0 & \lambda_N
\end{pmatrix}$.
\end{remark}

\paragraph{Assumptions and results }\label{FSARsec2} \mbox{} \\
Let $\gamma_{D^*} = (\alpha+1, B_{D^*}^\top)^\top$, $\gamma^* = (\alpha^*+1, B^{*\top})^\top$, $\Psi_{D^*} = (\sigma^2, \rho, \gamma_{D^*}^\top)^\top = (\Psi_1, \Psi_2, \dots, \Psi_{s_p(D^*)+2})^\top$, \\ $\Psi^* = (\sigma^{2*}, \rho^*, \gamma^{*\top})^\top = (\Psi_{1}^*, \Psi_{2}^*, \dots, \Psi_{s_p(D^*)+2}^*)^\top$, $G_N = V_N S_N^{-1}$, and \\
$Q_N(\Psi_{D^*}) = \ell_N(\Psi_{D^*}) - N \lambda_N B_{D^*}^\top B_{D^*}$. \\

Similarly to \cite{liu2018penalized}, we assume that (\textbf{Assumptions I}):
\begin{itemize}
\item[i.] Exists $\nu > 0$ such that for all $i \in \{ 1, \dots, N \}, \mathbb{E}[|\varepsilon_i|^{\nu + 4}]$ exists.
\item[ii.] $v_{ij,N}=O(h_N^{-1})$ uniformly in all $i,j$, where the rate sequence $h_N$ can be bounded or divergent, such as $h_N=o(N)$. 
\item[iii.] The matrix $S_N$ is nonsingular.
\item[iv.] The sequences of matrices $\{V_N\}$ and $\{S_N^{-1}\}$ are uniformly bounded in both row and column sums.
\item[v.] $\underset{N \rightarrow \infty}{\lim} \ \dfrac{1}{N} \chi_{N}^\top \chi_{N}$ exists and is nonsingular. The elements of $\chi_{N}$ are uniformly bounded constants.
\item[vi.] The matrices $\{S_N^{-1}(\rho)\}$ are uniformly bounded in both row and column sums and uniformly bounded in $\rho$ in compact parameter space $\mathcal{R}$. The true $\rho^*$ is in the interior of $\mathcal{R}$.
\item[vii.] $\underset{N \rightarrow \infty}{\lim} \ \dfrac{1}{N} [\chi_{N}, G_N \chi_{N} \gamma^*]^\top [\chi_{N}, G_N \chi_{N} \gamma^*]$ exists and is nonsingular.
\item[viii.] $\underset{N \rightarrow \infty}{\lim} \ \dfrac{1}{N} \mathbb{E}\left[ \dfrac{\partial^2 \ell_N(\Psi^*)}{\partial \Psi_{D^*} \partial \Psi_{D^*}^\top } \right]$ exists.
\item[ix.] The third derivatives $\dfrac{\partial^3 \ell_N(\Psi_{D^*})}{\partial \Psi_{j} \partial \Psi_{k} \partial \Psi_{l}}$ exist ($1 \le j,k,l \le s_p(D^*)+2$) for all $\Psi_{D^*}$ in an open set $\mathcal{B}$ that contains the true parameters $\Psi^*$. There are functions $M_{j,k,l}$ such that $\mathbb{E}[M_{j,k,l}] < \infty$ and $\left| \dfrac{1}{N} \dfrac{\partial^3 \ell_N(\Psi_{D^*})}{\partial \Psi_{j} \partial \Psi_{k} \partial \Psi_{l}} \right| \le M_{j,k,l}$ for all $\Psi_{D^*} \in \mathcal{B}$.
\end{itemize}
		
\begin{theorem} \label{th1}
Let $p_{\lambda_N}(\Psi_j) = \lambda_N \Psi_j^2$ if $4 \le j \le s_p(D^*)+2$, 0 otherwise. \\
Let $a_N = \underset{j \in \{ 4, \dots, s_p(D^*)+2 \}\rrbracket }{ \max } \ |p'_{\lambda_N}(\Psi_j^*)| = 2 \lambda_N \underset{j \in \{ 4, \dots, s_p(D^*)+2 \} }{ \max } \ |\Psi_j^*|$ \\ and $b_N = \underset{j \in \{ 4, \dots, s_p(D^*)+2 \} }{ \max } \ |p''_{\lambda_N}(\Psi_j^*)| = 2 \lambda_N$. \\
If 
$b_N = o(1)$ (i.e. $\underset{N \to \infty}{\lim} \ \lambda_N = 0$),
then under the \textbf{Assumptions I(i)-(ix)} there exists a local maximizer $\hat{\Psi}_{D^*}$ of $Q_N(\Psi_{D^*})$ such that
$$ || \hat{\Psi}_{D^*} - \Psi^* || = O_{\mathbb{P}}(N^{-1/2} + a_N).$$
\end{theorem}

\begin{theorem} \label{th2}
Let $p_{\lambda_N}(\Psi_j) = \lambda_N \Psi_j^2$ if $4 \le j \le s_p(D^*)+2$, 0 otherwise. \\
Let $a_N = \underset{j \in \{ 4, \dots, s_p(D^*)+2 \} }{ \max } \ |p'_{\lambda_N}(\Psi_j^*)| = 2 \lambda_N \underset{j \in \{ 4, \dots, s_p(D^*)+2 \} }{ \max } \ |\Psi_j^*|$ \\ and $b_N = \underset{j \in \{ 4, \dots, s_p(D^*)+2 \} }{ \max } \ |p''_{\lambda_N}(\Psi_j^*)| = 2 \lambda_N$. \\
If $\sqrt{N} a_N = O(1)$ and $b_N = o(1)$ (i.e. $\underset{N \to \infty}{\lim} \ \lambda_N = 0$), then under the \textbf{Assumptions I(i)-(ix)}

$$ \sqrt{N} [(\mathbb{E}(A)+P_2) (\hat{\Psi}_{D^*} - \Psi^*) + P_1] \overset{d}{\underset{N \rightarrow \infty}{\longrightarrow}} \mathcal{N}\left(0, \underset{N \rightarrow \infty}{\lim} \ \Sigma_N \right) $$
where $A = - \dfrac{1}{N} \dfrac{\partial^2 \ell_N(\Psi^*)}{\partial \Psi_{D^*} \partial \Psi_{D^*}^\top}$, $P_1 = \left(0,0,0,p'_{\lambda_N}(\Psi_{4}^*), \dots, p'_{\lambda_N}(\Psi_{s_p(D^*)+2}^*)\right)^\top$, \\ $P_2 = \text{diag}\left(0,0,0,p''_{\lambda_N}(\Psi_{4}^*), \dots, p''_{\lambda_N}(\Psi_{s_p(D^*)+2}^*)\right)$ and

\begin{small}
$$ \Sigma_N = \begin{pmatrix}
\dfrac{\mathbb{E}(\varepsilon_1^4)-\sigma^{4*}}{4 \sigma^{8*}} & \dfrac{\mathbb{E}(\varepsilon_1^4)-\sigma^{4*}}{2N \sigma^{6*}} \text{tr}(G_N) + \dfrac{\mathbb{E}(\varepsilon_1^3)}{2N\sigma^{6*}} \mathbf{1}_N^\top G_N \chi_{N} \gamma^* & \dfrac{\mathbb{E}(\varepsilon_1^3)}{2N\sigma^{6*}} \mathbf{1}_N^\top \chi_{N} \\
 & \dfrac{1}{N\sigma^{2*}} (G_N \chi_{N} \gamma^*)^\top (G_N \chi_{N} \gamma^*) & \\
* & + \dfrac{\mathbb{E}(\varepsilon_1^4)-3\sigma^{4*}}{N \sigma^{4*}} \dsum_{i=1}^N G_{ii,N}^2 + \dfrac{1}{N} \text{tr}((G_N^\top + G_N) G_N) & \dfrac{1}{N \sigma^{2*}} \gamma^{*\top} \chi_{N}^\top G_N^\top \chi_{N} + \dfrac{\mathbb{E}(\varepsilon_1^3)}{N\sigma^{4*}} \dsum_{i=1}^N G_{ii,N} \chi_{i.,N} \\
& + 2 \dfrac{\mathbb{E}(\varepsilon_1^3)}{N\sigma^{4*}} \dsum_{i=1}^N G_{ii,N} G_{i.,N} \chi_{N} \gamma^* &  \\
* & * & \dfrac{1}{N\sigma^{2*}} \chi_{N}^\top \chi_{N} \\
\end{pmatrix}.$$
\end{small}
\end{theorem}

\subsubsection{Spatial autoregressive model based on signatures projections (ProjSSAR)}
In this section we consider Model (\ref{modelsignaturedevelop}). Using the notation $B_{\infty}^* = (\beta_1^*, \dots, \beta_{(i_1, \dots, i_d)}^*, \dots)^\top $, 
and $\xi_{N,\infty}$ the matrix whose $i^\text{th}$ line corresponds to $\mathcal{S}(X_i)$ without the constant 1, one can rewrite (\ref{modelsignaturedevelop}) as 
\begin{equation*}
S_N\mathbf{Y}_N= (\alpha^*+1) \mathbf{1}_N + \xi_{N,\infty} B_{\infty}^*+\bm{\varepsilon}_N \, , \qquad N=1,2,\ldots
\end{equation*}

Then, the associated conditional quasi log-likelihood function of the vector $\mathbf{Y}_N$ given $\{Sig(X_i), \; i=1,\ldots,N\}$ is given by:
\begin{align}
\ell_N(\sigma^2,\rho,\alpha+1, B_{\infty})&= -\frac{N}{2}\ln(\sigma^2)-\frac{N}{2}\ln(2\pi)+ \ln\vert S_N(\rho)\vert \nonumber \\
&\hspace{0.5cm} -\frac{1}{2\sigma^2}\left[S_N(\rho)\mathbf{Y}_N - (\alpha + 1) \mathbf{1}_N- \xi_{N,\infty} B_{\infty}\right]^\top \left[S_N(\rho)\mathbf{Y}_N - (\alpha + 1) \mathbf{1}_N- \xi_{N,\infty} B_{\infty}\right]
\label{LK}.
\end{align}

\paragraph{Estimation} \mbox{} \\
We assume that there exist new coefficients $ \bm{\zeta}_i = (\zeta_{i,1}, \zeta_{i,2}, \dots )^\top $ and $ \Phi^* = (\phi_{1}^*, \phi_{2}^*, \dots )^\top $ such that $$\langle \theta^*, Sig(X_i) \rangle = \mathcal{S}(X_i) \left(1, B_{\infty}^{*\top}\right)^\top = 1 + \langle \bm{\zeta}_i, \Phi^* \rangle, $$ and a positive sequence of integers $C_N$ increasing asymptotically as the sample size $N \rightarrow \infty$ such that
$$ \langle \bm{\zeta}_i, \Phi^* \rangle = \dsum_{c=1}^{C_N} \zeta_{i,c} \phi_c^* + \dsum_{c=C_N+1}^{\infty} \zeta_{i,c} \phi_c^*$$ where the second term vanishes asymptotically when $N \rightarrow \infty$. \\

In practice the new coefficients $\bm{\zeta}_i$ can be computed by the use of a principal component analysis (PCA) on the signatures coefficients derived up to a pre-defined large truncation order, or by using another projection method.  \\

Then $\xi_{N,\infty} B_{\infty}^*$ can be approximated by $Z_{C_N} \Phi_{C_N}^*$ where $Z_{C_N}$ is the $N \times C_N$ matrix whose $i^\text{th}$ line is given by $$\bm{\zeta}_i^{C_N\top} = (\zeta_{i,1}, \dots, \zeta_{i,C_N}),$$ and $\Phi_{C_N}^* = (\phi_{1}^*, \dots, \phi_{C_N}^* )^\top$. \\

We suppose that the $\zeta_i^{C_N}$ are centered (which can be easily satisfied by subtracting their average) and that $\alpha^*+1$=0 which can be obtained by centering $\mathbf{Y}_N$.
Now, the truncated and feasible conditional quasi log-likelihood function can be obtained by replacing in (\ref{LK}) $\xi_{N,\infty} B_{\infty}$ with $Z_{C_N} \Phi_{C_N}$: 
\begin{eqnarray}
\tilde{\ell}_N(\sigma^2,\rho,\Phi_{C_N})&=&-\frac{N}{2}\ln(\sigma^2)-\frac{N}{2}\ln(2\pi)+  \ln\vert S_N(\rho)\vert \nonumber\\ && -\frac{1}{2\sigma^2}\left[S_N(\rho)\mathbf{Y}_N- Z_{C_N} \Phi_{C_N} \right]^\top \left[S_N(\rho)\mathbf{Y}_N- Z_{C_N} \Phi_{C_N} \right].
\label{TLKACP}
\end{eqnarray}

For a fixed $\rho$, (\ref{TLKACP}) is maximized at 
\begin{equation*}
\hat{\Phi}_{N,{\rho}}=(Z_{C_N}^\top Z_{C_N})^{-1}Z_{C_N}^\top S_N(\rho)\mathbf{Y}_N
\end{equation*} 
and 
\begin{eqnarray*}
\hat{\sigma}_{N,{\rho}}^2&=&\frac{1}{N}\left(S_N(\rho)\mathbf{Y}_N-Z_{C_N}\hat{\Phi}_{N,\rho}\right)^\top\left(S_N(\rho)\mathbf{Y}_N-Z_{C_N}\hat{\Phi}_{N,{\rho}}\right)\nonumber \\
&=&\frac{1}{N}\mathbf{Y}_N^\top S_N^\top(\rho)M_N S_N(\rho)\mathbf{Y}_N,
\end{eqnarray*}
where $M_N=I_N-Z_{C_N}(Z_{C_N}^\top Z_{C_N})^{-1}Z_{C_N}^\top$.\\

Thus, the concentrated truncated conditional quasi log-likelihood function of $\rho$ is:
\begin{equation*}
\tilde{\ell}_N(\rho)=-\frac{N}{2}[\ln(2\pi)+1] -\frac{N}{2}\ln(\hat{\sigma}^2_{N,\rho}) \, +\ln\vert S_N(\rho)\vert. 
\end{equation*} 
The estimator of $\rho^*$ is then $\hat{\rho}_N = \underset{\rho \in \mathcal{R}}{\arg \max} \ \tilde{\ell}_N(\rho)$, and those of $\Phi_{C_N}^{*}$ and $\sigma^{2*}$ are respectively $\hat{\Phi}_{N,\hat{\rho}_N}$ and $\hat{\sigma}^2_{N,\hat{\rho}_N}$.

\paragraph{Assumptions and results} \mbox{} \\
Let $I_N+\rho^* G_N=S_N^{-1}$, $K_N(\rho)=S_N(\rho)S_N^{-1}=I_N+(\rho^*-\rho)G_N$ for all $\rho\in \mathcal{R}$, $A_N(\rho)=K_N^\top(\rho) K_N(\rho)$ and $\Gamma_{C_N}=\mathbb{E}\left[ \dfrac{1}{N} Z_{C_N}^\top Z_{C_N} \right]$.\\
We assume that (\textbf{Assumptions II}):
\begin{itemize}
\item[i.] Assumptions I(i)-(iv) hold.
\item[ii.] Assumption I(vi) holds.
\item[iii.] The sequence $C_N$	satisfies $C_N\to \infty $ and  $C_N N^{-1/4}\to 0$ as $N\to \infty$, and 
\begin{align*}
&\mbox{1.}\quad C_N \sum_{r_1,r_2>C_N}\mathbb{E}\left(\zeta_{r_1}\zeta_{r_2}\right)=o(1)\\
&\mbox{2.}\quad \sum_{r_1,\ldots,r_4>C_N}\mathbb{E}\left(\zeta_{r_1} \ldots \zeta_{r_4}\right)=o(1)\\
&\mbox{3.}\qquad \sqrt{N}\sum_{s=1}^{C_N}\sum_{r_1,r_2>C_N}\mathbb{E}\left(\zeta_s \zeta_{r_1}\right)\mathbb{E}\left(\zeta_s \zeta_{r_2}\right)=o(1) \\
&\mbox{4.}\qquad
\sum_{r=1}^{C_N}\mathbb{E}\left( \zeta_{r}^{2}\right) < \infty.
\end{align*}
where $\zeta_{r}$ denotes the $r^\text{th}$ element of $\bm{\zeta}$.
\item[iv.] Let $\Delta_N = N\left(\text{tr}\left(\frac{G_N^\top G_N}{N}\right) - \text{tr}^{2}\left(\frac{G_N}{N}\right)\right)\Phi_{C_N}^{*\top}\Gamma_{C_N}\Phi_{C_N}^{*}$ and $\lim_{N\to\infty }\frac{1}{N}\Delta_N=c$, where (i) $c> 0$; (ii) $c=0$. \\
Under the latter  condition, 
	\begin{equation*}
	\lim_{N\to \infty} \frac{h_N}{N}\left\{\ln \left\vert \sigma^{2*}S_N^{-1}S_N^{-\top}\right\vert -\ln\left\vert \sigma_{N,\rho}^{2}S_N^{-1}(\rho)S_N^{-\top}(\rho)\right\vert\right\}\neq 0,
	\end{equation*} 
	whenever $\rho\neq \rho^*,$ with $ \sigma^2_{N,\rho}=\frac{\sigma^{2*}}{N}\mathrm{tr}(A_N(\rho))$.
\item[v.] We have
\begin{equation*}
\sum_{r_1,r_2,r_3,r_4=0}^{C_N}\mathbb{E}\left( \zeta_{r_1} \zeta_{r_2} \zeta_{r_3} \zeta_{r_4}\right)\nu_{r_1 r_2}\nu_{r_3 r_4}=o\left(\dfrac{N}{C_N^{2}}\right),
\end{equation*}
where the $\nu_{kl},\; k,l=1,\ldots, C_N$, are the elements of $ \Gamma_{C_N}^{-1}$.
\item[vi.] We assume that
\begin{equation*}
\sum_{r_1,\ldots,r_8=0}^{C_N}\mathbb{E}\left( \zeta_{r_1} \zeta_{r_3} \zeta_{r_5} \zeta_{r_7}\right)\\
\mathbb{E}\left( \zeta_{r_2} \zeta_{r_4} \zeta_{r_6}
\zeta_{r_8}\right) \nu_{r_1r_2}\nu_{r_3r_4}\nu_{r_5r_6}\nu_{r_7r_8}=o(N^2 C_N^{2}).
\end{equation*}
\end{itemize}

Then, similarly to \cite{ahmed2022quasi}, the following theorems give the identification, consistency and asymptotic normality results of the parameters estimates. 
\begin{theorem}\label{FSARth1}
Under \textbf{Assumptions II(i)-(iv)}  and $h_N^4=O(N)$ for divergent $h_N$, the QMLE $\hat{\rho}_N$ derived from the maximization of $\tilde{\ell}_N(\rho)$ is consistent and satisfies 
\begin{equation*}
\sqrt{\frac{N}{h_N}}(\hat{\rho}_N-\rho^*)\overset{d}{\underset{N \rightarrow \infty}{\longrightarrow}} \mathcal{N}(0,s_{\rho}^{2}),
\end{equation*}
with 	
$\displaystyle  s_{\rho}^{2}=\lim_{N\to \infty}\frac{s_{N}^2h_N}{N}\left\{\frac{h_N}{N}\left[\Delta_N+\sigma^{2*}\text{tr}((G^\top_N+G_N)G_N)\right]\right\}^{-2},$ where
\begin{eqnarray}
s_{N}^2&=& (\mathbb{E}(\varepsilon_1^4) - 2\sigma^{4*})\sum_{i=1}^N G_{ii}^2 + \frac{1}{N} \text{tr}^2(G_N) (\sigma^{4*}-\mathbb{E}(\varepsilon_1^4)-\sigma^{2*} \Phi_{C_N}^{*\top} \Gamma_{C_N} \Phi_{C_N}^*)  \nonumber \\ 
&& \qquad\qquad + \text{tr}(G_NG_N^\top)(\sigma^{4*}+\sigma^{2*} \Phi_{C_N}^{*\top} \Gamma_{C_N} \Phi_{C_N}^*).
\label{sn}
\end{eqnarray}
\end{theorem}

\begin{theorem}\label{FSARth3}
Under assumptions of Theorem~\ref{FSARth1}, $\hat{\sigma}_{N}^{2}$ is a consistent estimator of $\sigma^{2*}$ and satisfies 
\begin{equation*}
\sqrt{N}(\hat{\sigma}_{N,{\hat{\rho}_N}}^2-\sigma^{2*})\overset{d}{\underset{N \rightarrow \infty}{\longrightarrow}} \mathcal{N}(0,s^{2}_{\sigma}),
\end{equation*}
with 
\begin{equation*}
s^{2}_{\sigma}=\mathbb{E}(\varepsilon_1^4)-\sigma^{4*}+4s_{\rho}^2\lim_{N \to \infty}h_N\left[\frac{\text{tr}(G_N)}{N}\right]^{2}.
\end{equation*}  
\end{theorem} 
\noindent When $h_N$  is divergent,  $s_{\sigma}^{2}$ will be reduced to $\mathbb{E}(\varepsilon_1^4)-\sigma^{4*}$. \\

\begin{theorem}\label{FSARth2}
Under \textbf{Assumptions II(i)-(vi)}, we have
\begin{equation*}
\frac{N\left(\hat{\Phi}_{N,\hat{\rho}_N}-\Phi_{C_N}^{*}\right)^\top \Gamma_{C_N}\left(\hat{\Phi}_{N,\hat{\rho}_N}-\Phi_{C_N}^{*}\right)-\sigma^{2*} C_N}{\sqrt{2C_N}} \overset{d}{\underset{N \rightarrow \infty}{\longrightarrow}}\mathcal{N}(0,\sigma^{4*}).
\end{equation*}
\end{theorem}

\section{Finite sample properties} \label{sec:simu}

A simulation study was then conducted to compare the performances of the proposed signatures-based spatial autoregressive model considering the penalized spatial regression (PenSSAR) and the signatures projections (ProjSSAR) strategies. We also compared them with the functional spatial autoregressive model proposed by \cite{ahmed2022quasi} (FSARLM).

\subsection{Design of the simulation study}

We considered a grid with $60 \times 60$ locations, where we randomly allocate N=200 spatial units.
Then we considered data generated according to the following two models, which respectively favor the FSARLM approach and our approaches presented in this article (the PenSSAR and the ProjSSAR) where \\
$$X_i(t) = (X_{i,1}(t), \dots, X_{i,p}(t))^\top, X_{i,k}(t) = \alpha_{i,k} t + f_{i,k}(t),$$
$$\theta^*(t) = (\theta_1^*(t), \dots, \theta_p^*(t))^\top, \theta^*_k(t) = \Psi_k t + g_{i,k}(t),$$ 
$$\alpha_{i,k} \sim \mathcal{U}([-3,3]) \text{ and } \Psi_k \sim \mathcal{U}([-3,3]).$$
$f_{i,k}$ and $g_{i,k}$ are Gaussian processes with exponential covariance matrix with length-scale 1, and $\varepsilon_i \sim \mathcal{N}(0,1)$ for $i \in \{ 1, \dots, N \}$.
$X_i$ is observed at 101 equally spaced times of $[0,1]$.
\begin{itemize}[leftmargin=1.9cm]
\item[Model 1.] $Y_i = \rho^* \dsum_{j=1}^N v_{ij,N} Y_j + \int_0^1 X_i(t)^\top \theta^*(t) \ \text{d}t + \varepsilon_i$.
\item[Model 2.] $Y_i = \rho^* \dsum_{j=1}^N v_{ij,N} Y_j + \langle \mathcal{S}^{D^*}(X_i), \mathcal{S}^{D^*}(\theta^*) \rangle + \varepsilon_i$
where $D^* = 2$.
\end{itemize}
For these models, we aim at predicting the $Y_i$ given the 101 observations of $X_i$ over the time. \\

Then, similarly to \cite{fermanian2022functional}, we investigated three other models which, in their design, do not correspond specifically to any of the approaches.
\begin{itemize}[leftmargin=1.9cm]
\item[Model 3.] $Y_i = \rho^* \dsum_{j=1}^N v_{ij,N} Y_j + ||\alpha_i|| + \varepsilon_i$, $\alpha_i = (\alpha_{i,1}, \dots, \alpha_{i,p})^\top $
\item[Model 4.] $Y_i = \rho^* \dsum_{j=1}^N v_{ij,N} Y_j + \dfrac{1}{p} \dsum_{k=1}^p X_{i,k}(t_{101}) + \varepsilon_i$
\item[Model 5.] $Y_i = \rho^* \dsum_{j=1}^N v_{ij,N} Y_j + \dfrac{1}{p} \dsum_{k=1}^p Z_{i,k}(t_{101}) + \varepsilon_i$
\end{itemize}
where $Z_i(t) =  (Z_{i,1}(t), \dots, Z_{i,p}(t))^\top, Z_{i,k}(t) = \beta_{i,k,1} + 10 \beta_{i,k,2} \sin\left( \dfrac{2 \pi t}{\beta_{i,k,3}} \right) + 10 (t - \beta_{i,k,4})^3$, $\beta_{i,k,j} \sim \mathcal{U}([0,1]), j = 1,2,3,4$. \\

For Model 3, we also aim at predicting the $Y_i$ given the observations of $X_i$. However in Models 4 and 5, we predict $Y_i$, which depends on the value of its neighbors and of the average of the components of $X_i$ (or $Z_i$) at time 101, from the observations of $X_i$ (or $Z_i$) at times 1 to 100. \\

The spatial weight matrix $V_N$ was constructed using the $k$-nearest neighbors method, and we considered the cases $k=4$ and $k=8$, $p=2,6,10$ and $\rho^*=0,0.2,0.4,0.6,0.8$. \\

Moreover, it should be noted that the signatures present the property to be invariant by translation and by time reparametrization \citep{lyons2007differential}. Thus, to circumvent the invariance by translation we added an observation point taking the value 0 at the beginning of $X_i$ ($Z_i$), and to avoid the invariance by time reparametrization we considered $\tilde{X}_i(t) = (X_i(t)^\top, t)$ ($\tilde{Z}_i(t) = (Z_i(t)^\top, t)$) before computing the signature of $X_i$ ($Z_i$).

For each model, several approaches were compared:

\begin{itemize}
\item[(i)] The approach proposed by \cite{ahmed2022quasi} (FSARLM) using a cubic B-splines basis with 12 equally spaced knots to approximate the $X_i$ ($Z_i$) from the observed data and a functional PCA \citep{ramsay2005functional}. As proposed by \cite{ahmed2022quasi}, we used a threshold on the number of coefficients such that the cumulative inertia was below $95\%$. 
\item[(ii)] Our proposed approach based on the penalized spatial regression (PenSSAR) with a truncation order $D^{\text{max}}$ such that there are at most $10^4$ signatures coefficients..
\item[(iii)] Our proposed approach based on signatures projections (ProjSSAR). We considered a maximum truncation order for the signatures to reach a maximal number of coefficients of $10^4$. Then a PCA was performed on the standardized truncated signature coefficients vectors, and similarly to \cite{ahmed2022quasi}, we used a threshold on the maximal number of coefficients such that the cumulative inertia was below $95\%$.
\end{itemize}

For each model and each value of $k,p$ and $\rho^*$, 200 data sets were generated. Each data set was then split into a training, a validation and a test set such that the optimal number of coefficients (for the FSARLM), the optimal truncation order (for the PenSSAR), and the optimal number of coefficients associated with the optimal truncation order (for the ProjSSAR) was selected on the validation set based on the root mean squared error (RMSE) criterion and the performances were finally evaluated on the test set using the RMSE. For generating the training, validation and test sets, we considered ordinary cross-validation and spatial cross-validation. For the latter we used a $K$-means algorithm (with $K=6$) on the coordinates of the data and two clusters were randomly selected to be the validation and test sets.

\subsection{Results of the simulation study}

\begin{figure}
\centering
\includegraphics[width=\textwidth]{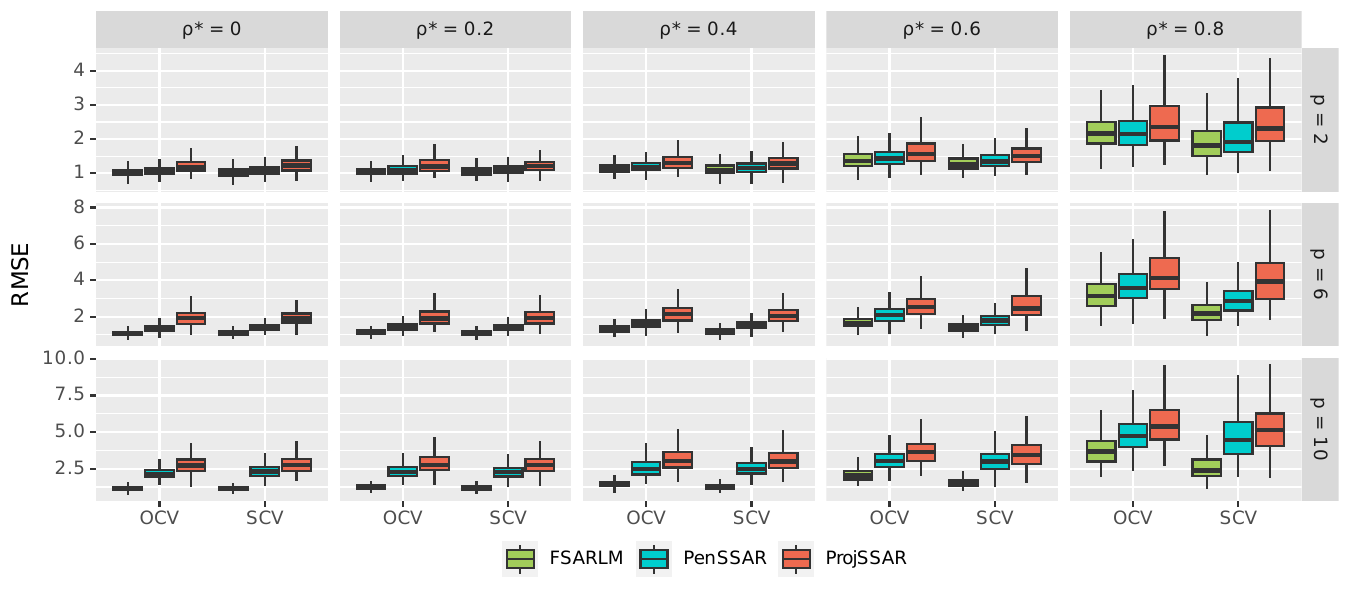}
\caption{RMSE on the test set with the FSARLM, the PenSSAR and the ProjSSAR for Model 1 with $k=4$ using ordinary (OCV) and spatial (SCV) cross-validation}
\label{fig:rmse4_1}
\end{figure}

\begin{figure}
\centering
\includegraphics[width=\textwidth]{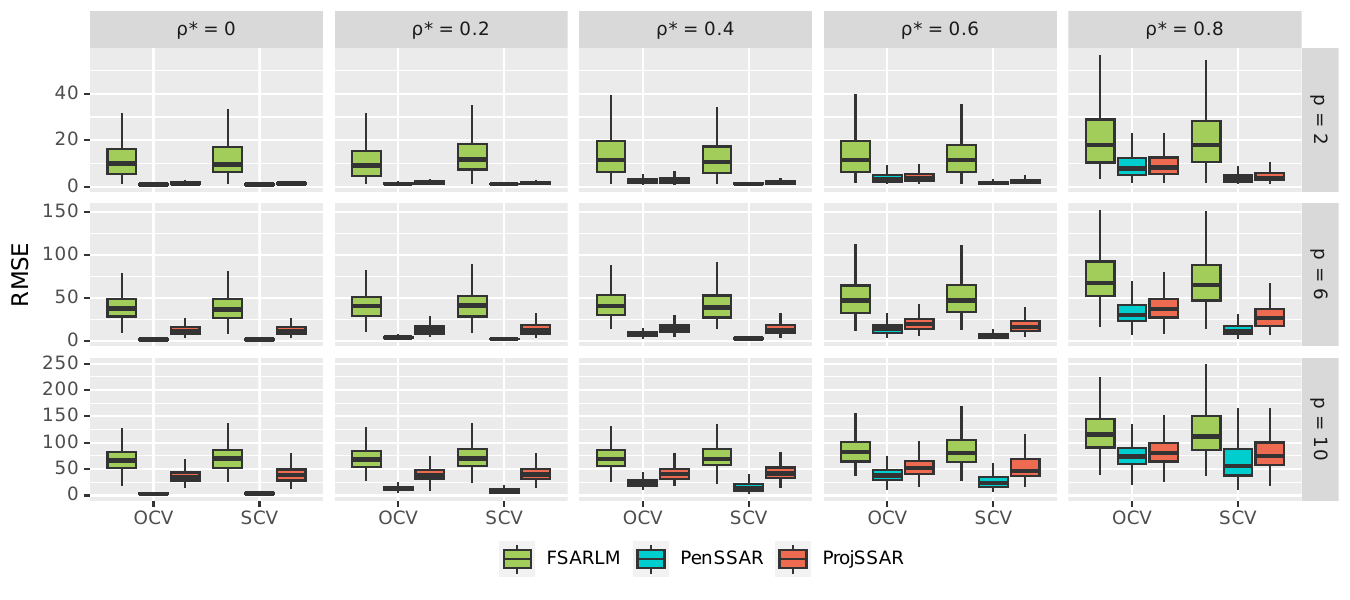}
\caption{RMSE on the test set with the FSARLM, the PenSSAR and the ProjSSAR for Model 2 with $k=4$ using ordinary (OCV) and spatial (SCV) cross-validation}
\label{fig:rmse4_2}
\end{figure}

\begin{figure}
\centering
\includegraphics[width=\textwidth]{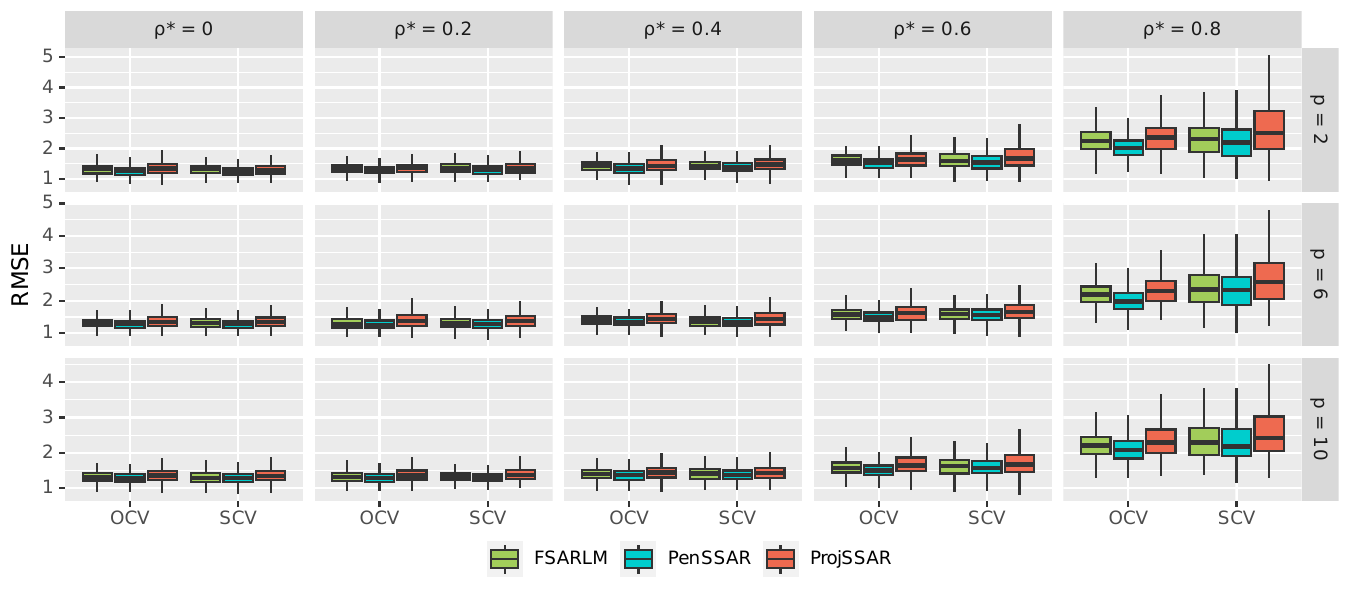}
\caption{RMSE on the test set with the FSARLM, the PenSSAR and the ProjSSAR for Model 3 with $k=4$ using ordinary (OCV) and spatial (SCV) cross-validation}
\label{fig:rmse4_3}
\end{figure}

\begin{figure}
\centering
\includegraphics[width=\textwidth]{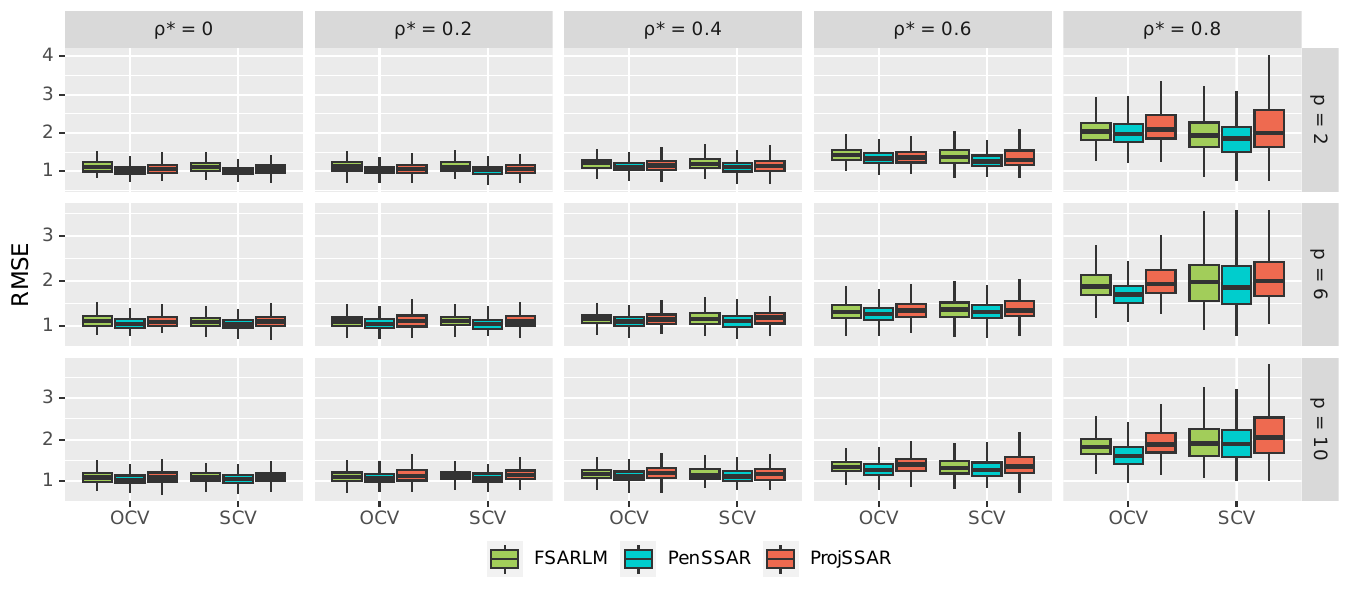}
\caption{RMSE on the test set with the FSARLM, the PenSSAR and the ProjSSAR for Model 4 with $k=4$ using ordinary (OCV) and spatial (SCV) cross-validation}
\label{fig:rmse4_4}
\end{figure}

\begin{figure}
\centering
\includegraphics[width=\textwidth]{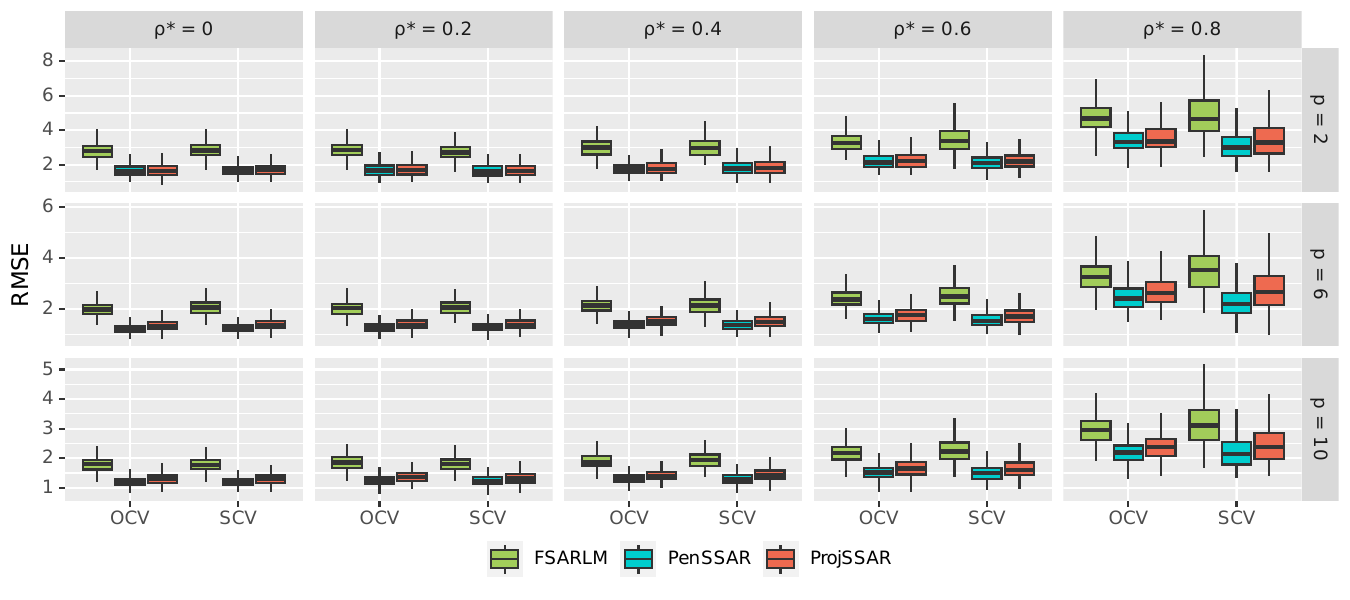}
\caption{RMSE on the test set with the FSARLM, the PenSSAR and the ProjSSAR for Model 5 with $k=4$ using ordinary (OCV) and spatial (SCV) cross-validation}
\label{fig:rmse4_5}
\end{figure}

For the sake of brevity, the results of the simulation study are presented in Figures \ref{fig:rmse4_1} to  \ref{fig:rmse4_5} for $k=4$ only. However it should be noted that similar results were obtained for $k=8$ (see the Appendix \ref{sec:k8} for the results).

For all models, the PenSSAR presents better performances than the ProjSSAR approach.

With Model 1, which is naturally favorable to the FSARLM, the latter presents the best performances. It should be noted however, that our PenSSAR approach presents close RMSEs, especially when $p=2$.

In the case of Model 2 (which is naturally favorable to our methods), the PenSSAR and the ProjSSAR present much lower RMSEs than the FSARLM whatever the values of $\rho^*$ or $p$.

When considering Models 3 and 4, all approaches present similar performances although our PenSSAR approach presents slightly lower RMSEs.

Finally, when considering Model 5, our PenSSAR and ProjSSAR approaches present much lower RMSEs than the FSARLM, whatever the values of $\rho^*$ and $p$. It should be noted that in this model, the process $Z$ is very irregular. Thus we tried to increase the number of knots when smoothing $Z_i$ for the FSARLM. However this did not improve the performances of the approach.

\section{Real data application} \label{sec:appli}

\begin{figure}
\centering
\begin{minipage}{0.49\linewidth}
\includegraphics[width=\linewidth]{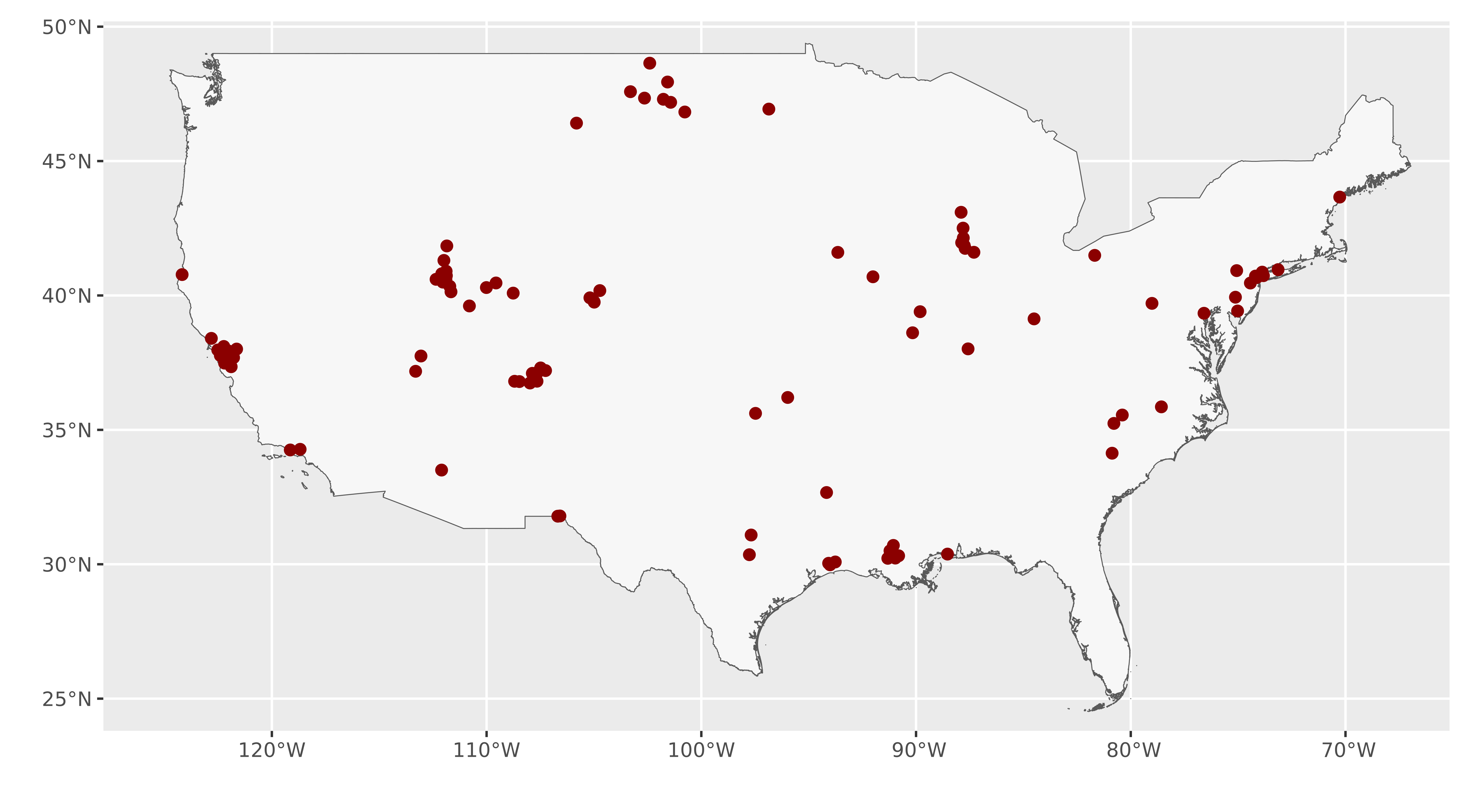}
\end{minipage} \hfill
\begin{minipage}{0.49\linewidth}
\includegraphics[width=\linewidth]{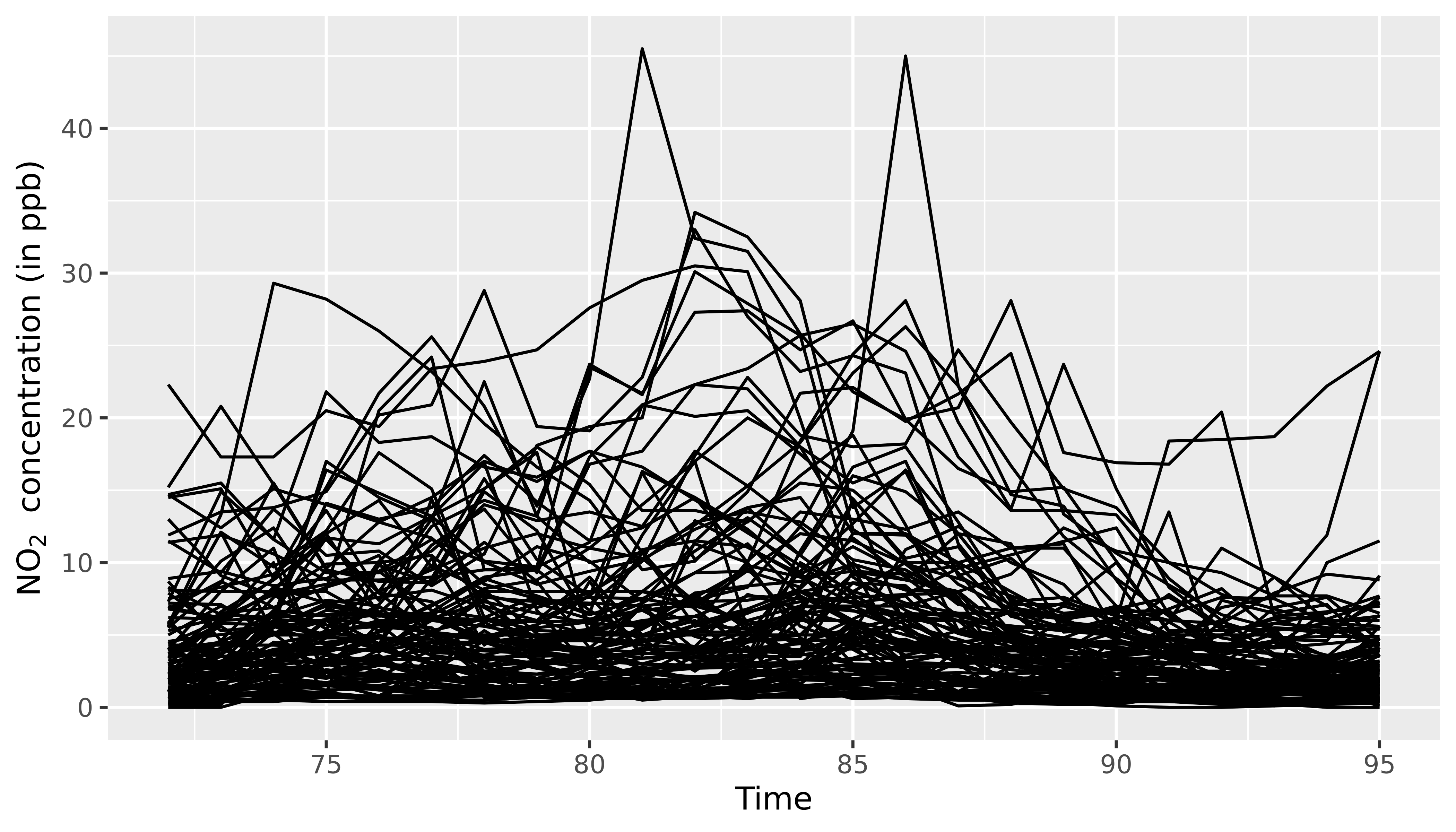}
\includegraphics[width=\linewidth]{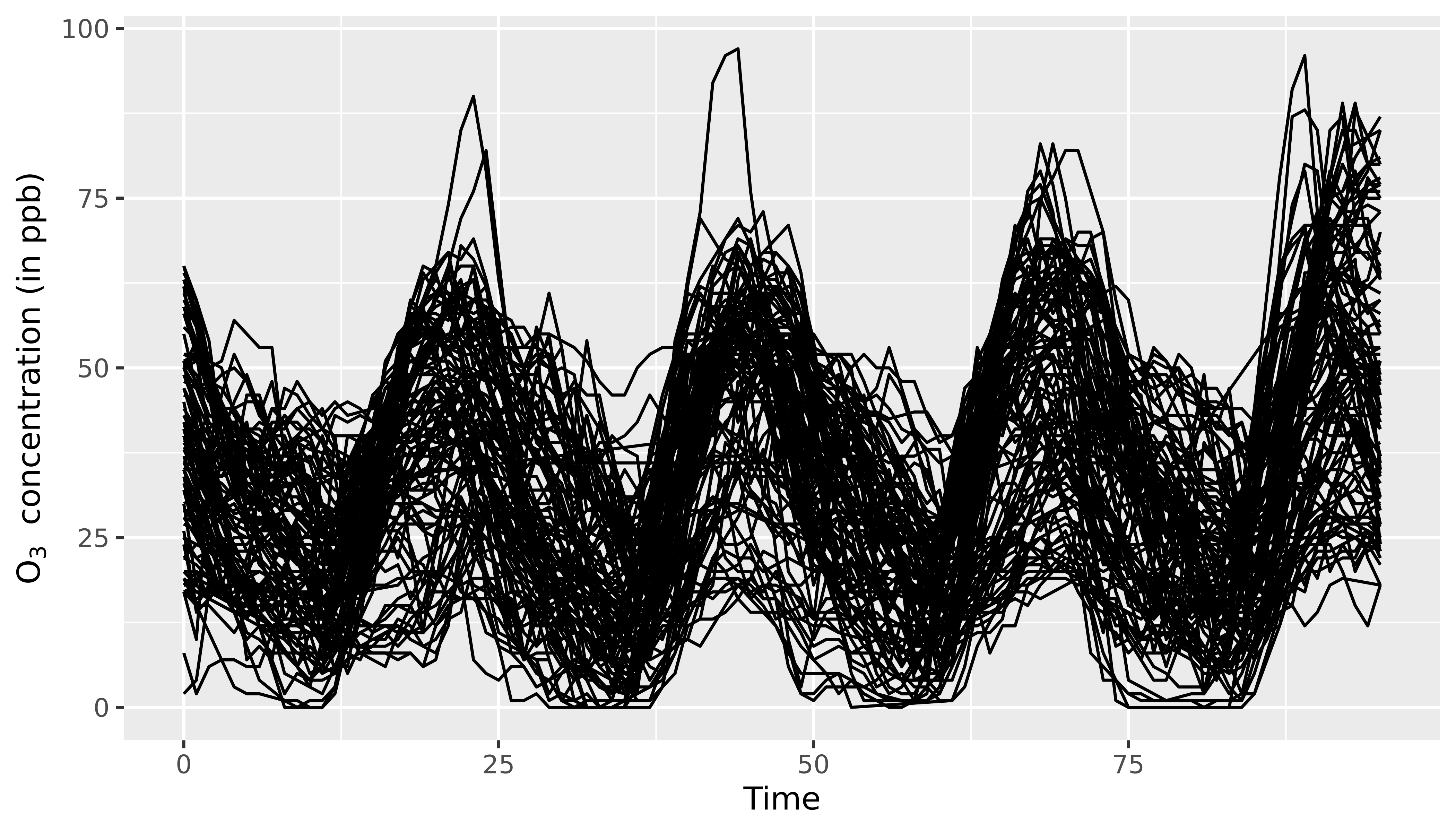}
\end{minipage}
\caption{Spatial locations of the 104 monitoring stations across the United States (left panel) and hourly nitrogen dioxide and ozone concentrations (from August 1, 2022, 0:00 to August 4, 2022, 23:00, and from August 4, 2022, 0:00 to August 4, 2022, 23:00, respectively)}
\label{fig:appli}
\end{figure}

In this section, we consider air quality data collected from 104 monitoring stations across the United States (\url{https://www.epa.gov/outdoor-air-quality-data}). The data consist in hourly ozone concentrations (in ppb) from August 1, 2022, 0:00 to August 4, 2022, 23:00, and hourly nitrogen dioxide concentrations (in ppb) from August 4, 2022, 0:00 to August 4, 2022, 23:00. We used linear interpolation to estimate the
missing values. The spatial locations of the monitoring stations, as well as the ozone and nitrogen dioxide concentrations are presented in Figure \ref{fig:appli}. \\

In this application, we aim at predicting
\begin{itemize}
    \item[(i)] the concentrations of nitrogen dioxide ($Y_i$) at each hour $H$:00 of August 4, 2022 from the ozone concentrations from August 1, 2022, 0:00 to August 4, 2022, $(H-1)$:00 ($\{X_i(t), t \in [0, 71+H] \}$),
    \item[(ii)] the average concentration of nitrogen dioxide ($Y_i$) on August 4, 2022 from the ozone concentrations from August 1, 2022, 0:00 to August 3, 2022, 23:00 ($\{X_i(t), t \in [0, 71] \}$), and
    \item[(iii)] the maximum concentration of nitrogen dioxide ($Y_i$) on August 4, 2022 from the ozone concentrations from August 1, 2022, 0:00 to August 3, 2022, 23:00 ($\{X_i(t), t \in [0, 71] \}$). 
\end{itemize}
To this end, we used the three approaches (FSARLM, PenSSAR and ProjSSAR), considering the following spatial weight matrix $V_N$, similarly to \cite{ahmed2022quasi}:
\begin{equation*}
v_{ij,N} = \begin{cases}
\dfrac{1}{1 + d_{ij}} &\text{ if $d_{ij} < l$ and $i \neq j$}\\
0 &\text{otherwise}
\end{cases}
\end{equation*}
where $d_{ij}$ corresponds to the Great Circle distance between the monitoring stations $i$ and $j$, and $l$ is a threshold determined so that all monitoring stations have at least four neighbors. \\
It should be noted that, for the same reasons as in the simulation study, for the PenSSAR and the ProjSSAR approaches, we added an observation point taking the value 0 at the beginning of $X_i$, and we considered $\tilde{X}_i(t) = (X_i(t)^\top, t)$ before computing the signature of $X_i$. \\
We also considered ordinary cross-validation and spatial cross-validation using a $K$-means algorithm (with $K=6$) on the coordinates of the data, and we repeated the procedure on 50 different train/validation/test sets for ordinary cross-validation and 30 different sets for spatial cross-validation (thus covering all the possibilities for the latter). \\

\begin{figure}
    \centering
    \includegraphics[width=\linewidth]{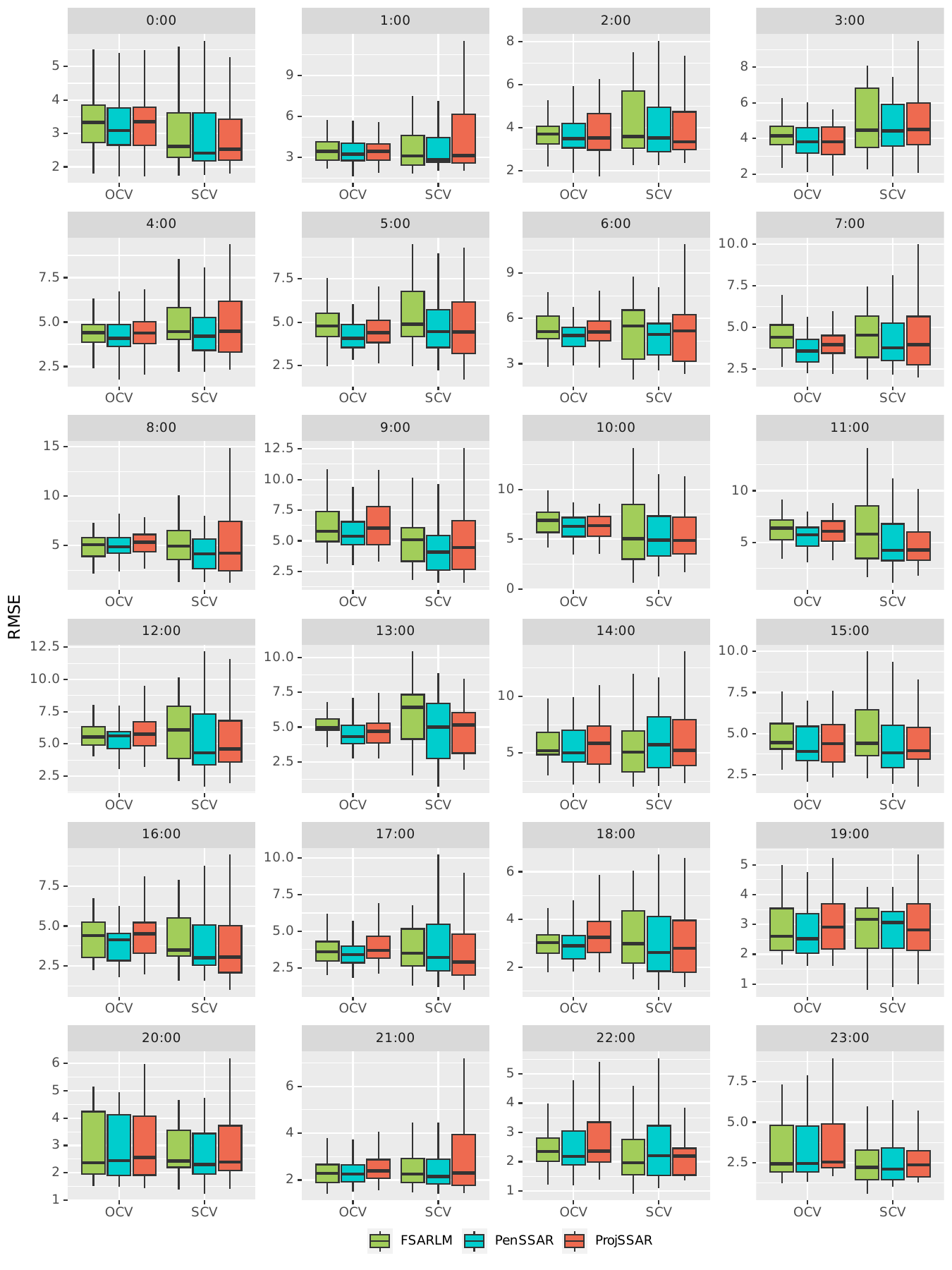}
    \caption{RMSE on the test set with the FSARLM, the PenSSAR and the ProjSSAR for predicting the concentrations of nitrogen dioxide at each hour $H$ of August 4, 2022 from the ozone concentrations from August 1, 2022, 0:00 to August 4, 2022, $(H-1)$:00, using ordinary (OCV) and spatial (SCV) cross-validation}
    \label{fig:appli1}
\end{figure}

\begin{figure}
    \centering
    \includegraphics[width=\linewidth]{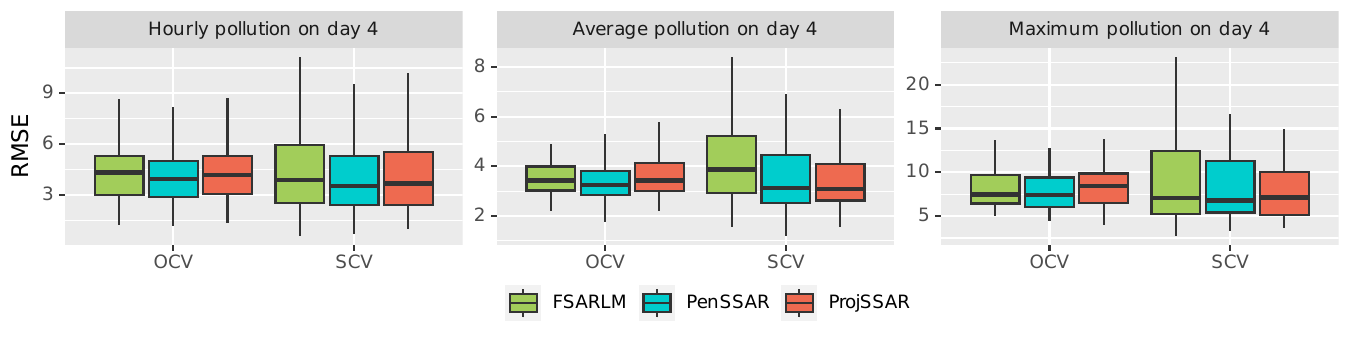}
    \caption{RMSE on the test set with the FSARLM, the PenSSAR and the ProjSSAR for predicting the concentrations of nitrogen dioxide at each hour $H$ of August 4, 2022 from the ozone concentrations from August 1, 2022, 0:00 to August 4, 2022, $(H-1)$:00 (left panel), and the average (middle panel) or the maximum (right panel) concentration of nitrogen dioxide on August 4, 2022 from the ozone concentrations from August 1, 2022, 0:00 to August 3, 2022, 23:00, using ordinary (OCV) and spatial (SCV) cross-validation}
    \label{fig:appli2}
\end{figure}

Figure \ref{fig:appli1} presents the hourly results for objective (i). It shows that for almost all 24 hours of prediction, the PenSSAR approach outperforms the FSARLM and the ProjSSAR method performs similarly to or better than the FSARLM. Figure \ref{fig:appli2} (panel ``Hourly pollution on day 4'') shows these performances for the whole day of August 4, 2022. For objectives (ii) and (iii), Figure \ref{fig:appli2} shows that when using an ordinary cross-validation, the PenSSAR approach performs slightly better than the FSARLM method, and the ProjSSAR performs similarly to the FSARLM. However, when using a spatial cross-validation, our two approaches perform better than the FSARLM, especially for predicting the average concentration (objective (ii)). 

\section{Discussion} \label{sec:conclu}
Here we proposed an alternative to the traditional spatial autoregressive model with functional covariates \citep{huang2018spatial, pineda2019functional, ahmed2022quasi}. This new approach is based on the notion of signatures and presents the advantages of being applicable to a wide range of processes that are not necessarily square-integrable processes, and to better capture the differences between the curves. We then proposed two methods for estimating the model, respectively based on a penalized regression (similarly to \cite{fermanian2022functional}, PenSSAR) and on signatures projections (using a PCA for example, ProjSSAR). We also provided their theoretical guarantees. \\

In a simulation study, when simulating the data according to the model of \cite{ahmed2022quasi} (FSARLM), the latter presented the best performances. However it should be noted that the PenSSAR presented close performances. When simulating the data according to the model considered here, the ProjSSAR and the PenSSAR perform much better than the FSARLM. Finally, we considered three other simulation models which do not correspond to the FSARLM model, nor to the one presented in this work. In this case, when considering rather regular covariates, all approaches present similar RMSEs although the PenSSAR approach presented slightly better performances, but when considering more irregular covariates, the ProjSSAR and the PenSSAR present much better performances than the FSARLM. \\

Finally, we applied the FSARLM, the PenSSAR and the ProjSSAR methods on a real data set corresponding to ozone and nitrogen dioxide concentrations measured in monitoring stations across the United States. When using an ordinary cross-validation, the PenSSAR approach outperforms the two other approaches while the ProjSSAR presented performances close to those of the FSARLM. However, when using a spatial cross-validation, the PenSSAR and the ProjSSAR outperform the FSARLM. \\

It should be noted that we have considered here only functional covariates. However, it would be interesting to be able to integrate both functional and non-functional covariates in the model. \\
Moreover the approach proposed here is designed to handle a univariate response variable. The case of a multivariate response variable should also be investigated. 
\cite{zhu2020multivariate} has recently proposed a spatial autoregressive model in this context with non-functional covariates. Thus, extending this approach to the context of functional covariates will be the subject of future work. \\

Another development could be the adaptation of our proposed approach to the  spatial error model (SEM) or the SARAR model (SAR model with SAR errors) considering spatially correlated errors, with functional covariates.

\bibliography{biblioFSAR.bib}
\bibliographystyle{plainnat}

\appendix

\begin{lemma} \label{lemme1}
Under \textbf{Assumptions I(i)-(vii)}, 
$$ \dfrac{1}{\sqrt{N}} \dfrac{\partial \ell_N(\Psi^*) }{\partial \Psi_{D^*}} = O_{\mathbb{P}}(1).$$
\end{lemma}

\section{Proof of Lemma \ref{lemme1}}

\underline{For $\sigma^2$ we have}:
\begin{align*}
\dfrac{\partial \ell_N(\Psi_{D^*})}{\partial \sigma^2} &= -\dfrac{N}{2\sigma^2} + \dfrac{1}{2 \sigma^4} [S_N(\rho) \mathbf{Y}_N - \chi_N \gamma_{D^*}]^\top [S_N(\rho) \mathbf{Y}_N - \chi_N \gamma_{D^*}]. 
\end{align*}
$\blacktriangleright$ \ In $\Psi_{D^*} = \Psi^*$ we obtain:
\begin{equation} \label{eq1}
\dfrac{1}{\sqrt{N}} \dfrac{\partial \ell_N(\Psi^*)}{\partial \sigma^2} = \dfrac{1}{2 \sigma^{4*} \sqrt{N}}[\bm{\varepsilon}_N^\top \bm{\varepsilon}_N - N \sigma^{2*}].
\end{equation}

\underline{For $\rho$ we have}:
\begin{eqnarray*}
\dfrac{\partial \ell_N(\Psi_{D^*})}{\partial \rho} &=& -\text{tr}(V_N S_N^{-1}(\rho)) \\ && - \dfrac{1}{2 \sigma^2} [-\mathbf{Y}_N^\top V_N^\top \mathbf{Y}_N - \mathbf{Y}_N^\top V_N \mathbf{Y}_N + 2\rho \mathbf{Y}_N^\top V_N^\top V_N \mathbf{Y}_N + \gamma_{D^*}^\top \chi_N^\top V_N \mathbf{Y}_N + \mathbf{Y}_N^\top V_N^\top \chi_N \gamma_{D^*}] .
\end{eqnarray*}
$\blacktriangleright$ \ In $\Psi_{D^*} = \Psi^*$ we obtain:
\begin{align*}
\dfrac{\partial \ell_N(\Psi^*)}{\partial \rho} &= - \text{tr}(G_N) + \dfrac{1}{2 \sigma^{2*}} [\mathbf{Y}_N^\top V_N^\top (I-\rho^* V_N) \mathbf{Y}_N + \mathbf{Y}_N^\top (I-\rho^* V_N^\top) V_N \mathbf{Y}_N - \gamma^{*\top} \chi_N^\top V_N \mathbf{Y}_N - \mathbf{Y}_N^\top V_N^\top \chi_N \gamma^*] \\
&= -\text{tr}(G_N) + \dfrac{1}{\sigma^{2*}} \bm{\varepsilon}_N^\top V_N \mathbf{Y}_N \\
&=  -\text{tr}(G_N) + \dfrac{1}{\sigma^{2*}} \bm{\varepsilon}_N^\top V_N S_N^{-1} \chi_N \gamma^* + \dfrac{1}{\sigma^{2*}} \bm{\varepsilon}_N^\top V_N S_N^{-1} \bm{\varepsilon}_N \\
&= -\text{tr}(G_N) + \dfrac{1}{\sigma^{2*}} (G_N \chi_N \gamma^*)^\top \bm{\varepsilon}_N + \dfrac{1}{\sigma^{2*}} \bm{\varepsilon}_N^\top G_N \bm{\varepsilon}_N . \\
\end{align*}
And so
\begin{equation} \label{eq2}
\dfrac{1}{\sqrt{N}} \dfrac{\partial \ell_N(\Psi^*)}{\partial \rho} = \dfrac{1}{\sqrt{N} \sigma^{2*}} (G_N \chi_N \gamma^*)^\top \bm{\varepsilon}_N + \dfrac{1}{\sqrt{N} \sigma^{2*}} [\bm{\varepsilon}_N^\top G_N \bm{\varepsilon}_N - \sigma^{2*} \text{tr}(G_N)].
\end{equation}

\underline{For $\gamma_{D^*}$ we have}:
\begin{align*}
\dfrac{\partial \ell_N(\Psi_{D^*})}{\partial \gamma_{D^*}} &= \dfrac{1}{\sigma^2} \chi_N^\top [S_N(\rho) \mathbf{Y}_N - \chi_N \gamma_{D^*}].
\end{align*}
$\blacktriangleright$ \ In $\Psi_{D^*} = \Psi^*$ we obtain:
\begin{equation} \label{eq3}
\dfrac{1}{\sqrt{N}} \dfrac{\partial \ell_N(\Psi^*)}{\partial \gamma_{D^*}} = \dfrac{1}{\sqrt{N} \sigma^{2*}} \chi_N^\top \bm{\varepsilon}_N .
\end{equation}

It should be noted that $\mathbb{E}\left[ \dfrac{1}{\sqrt{N}} \dfrac{\partial \ell_N(\Psi_{D^*}) }{\partial \sigma^2} \right] = \mathbb{E}\left[ \dfrac{1}{\sqrt{N}} \dfrac{\partial \ell_N(\Psi_{D^*}) }{\partial \rho} \right] = \mathbb{E}\left[ \dfrac{1}{\sqrt{N}} \dfrac{\partial \ell_N(\Psi_{D^*}) }{\partial \gamma_{D^*}} \right] = 0$. \\

Then we can compute the variances:

\begin{align*}
\mathbb{V}\left( \dfrac{1}{\sqrt{N}} \dfrac{\partial \ell_N(\Psi^*)}{\partial \sigma^2} \right) &= \mathbb{V}\left( \dfrac{1}{2 \sigma^{4*} \sqrt{N}}[\bm{\varepsilon}_N^\top \bm{\varepsilon}_N - N \sigma^{2*}] \right) \\
&= \dfrac{1}{4 \sigma^{8*} N} \mathbb{V}(\bm{\varepsilon}_N^\top \bm{\varepsilon}_N) \\
&= \dfrac{1}{4 \sigma^{8*}} \mathbb{V}(\varepsilon_1^2) \\
&= \dfrac{1}{4 \sigma^{8*}} [\mathbb{E}(\varepsilon_1^4) - \sigma^{4*}] = O(1) \text{ by Assumption I(i) .} \\
\end{align*}
Thus $ \dfrac{1}{\sqrt{N}} \dfrac{\partial \ell_N(\Psi^*)}{\partial \sigma^2} = O_{\mathbb{P}}(1)$ by Chebyshev's inequality. \\

\begin{align*}
\mathbb{V} \left( \dfrac{1}{\sqrt{N}} \dfrac{\partial \ell_N(\Psi^*)}{\partial \rho} \right) &= \mathbb{V} \left( \dfrac{1}{\sqrt{N} \sigma^{2*}} (G_N \chi_N \gamma^*)^\top \bm{\varepsilon}_N + \dfrac{1}{\sqrt{N} \sigma^{2*}} [\bm{\varepsilon}_N^\top G_N \bm{\varepsilon}_N - \sigma^{2*} \text{tr}(G_N)] \right) \\
&\le 2 \left[  \mathbb{V}\left( \dfrac{1}{\sqrt{N} \sigma^{2*}} (G_N \chi_N \gamma^*)^\top \bm{\varepsilon}_N \right) +  \mathbb{V}\left( \dfrac{1}{\sqrt{N} \sigma^{2*}} (\bm{\varepsilon}_N^\top G_N \bm{\varepsilon}_N - \sigma^{2*} \text{tr}(G_N)) \right) \right] \\
&= \dfrac{2}{N \sigma^{2*}} (G_N \chi_N \gamma^*)^\top (G_N \chi_N \gamma^*) + \dfrac{2}{N \sigma^{4*}} \mathbb{V}(\bm{\varepsilon}_N^\top G_N \bm{\varepsilon}_N) \\
&= O(1) + O\left( \dfrac{1}{h_N} \right) = O(1)
\end{align*}
 by using the inequality for two real random variables $R_1$ and $R_2$: $$\mathbb{V}(R_1 + R_2) \le 2[\mathbb{V}(R_1) + \mathbb{V}(R_2)].$$

Thus, $\dfrac{1}{\sqrt{N}} \dfrac{\partial \ell_N(\Psi^*)}{\partial \rho} = O_{\mathbb{P}}(1)$ by Chebyshev's inequality. \\

Finally 
\begin{align*}
\mathbb{V}\left(\dfrac{1}{\sqrt{N}} \dfrac{\partial \ell_N(\Psi^*)}{\partial \gamma_{D^*}} \right) &= \dfrac{1}{N \sigma^{4*}} \chi_N^\top \mathbb{V}(\bm{\varepsilon}_N) \chi_N \\
&= \dfrac{1}{N \sigma^{2*}} \chi_N^\top \chi_N = O(1) \text{ by Assumption I(v)}.
\end{align*}

Thus, $\dfrac{1}{\sqrt{N}} \dfrac{\partial \ell_N(\Psi^*)}{\partial \gamma_{D^*}} = O_{\mathbb{P}}(1)$ by Chebyshev's inequality.
\begin{flushright} $\blacksquare$ \end{flushright}

\begin{lemma} \label{lemme2}
 Under \textbf{Assumptions I(i)-(viii)} we have \citep{lee2004asymptotic}
    $$ \dfrac{1}{N} \dfrac{\partial^2 \ell_N(\Psi^*)}{\partial \Psi_{D^*} \partial \Psi_{D^*}^\top } = \mathbb{E}\left[ \dfrac{1}{N} \dfrac{\partial^2 \ell_N(\Psi^*)}{\partial \Psi_{D^*} \partial \Psi_{D^*}^\top } \right] + o_{\mathbb{P}}(1).$$
\end{lemma}

\section{Proof of Theorem \ref{th1}}

Let $\alpha_N = N^{-1/2} + a_N$.
It suffices to show that for all $\eta > 0$, there exists a sufficiently large constant $C$ such that
\begin{equation} \label{equation}
\mathbb{P}\left[ \underset{||u||=C}{\sup} \ Q_N(\Psi^* + \alpha_N u) < Q_N(\Psi^*)\right] \ge 1 - \eta .
\end{equation}

\begin{flalign*}
\mathcal{D}_N(u) &= Q_N(\Psi^* + \alpha_N u) - Q_N(\Psi^*) & \\
&= \ell_N(\Psi^* + \alpha_N u) - \ell_N(\Psi^*) - N \dsum_{j=4}^{s_p(D^*)+2} [p_{\lambda_N}(\Psi_{j}^* + \alpha_N u_j) - p_{\lambda_N}(\Psi_{j}^*)] . &
\end{flalign*}
By using a Taylor expansion at order 2 on $\ell_N$ and $p_{\lambda_N}$ we get:
\begin{eqnarray*}
\mathcal{D}_N(u) &=& \alpha_N \left( \dfrac{\partial \ell_N(\Psi^*)}{\partial \Psi_{D^*}} \right)^\top u + \dfrac{1}{2} \alpha_N^2 u^\top \dfrac{\partial^2 \ell_N(\Psi^*)}{\partial \Psi_{D^*} \partial \Psi_{D^*}^\top} u (1 + o_{\mathbb{P}}(1)) \\ 
&&- N \dsum_{j=4}^{s_p(D^*)+2} \left[ p'_{\lambda_N}(\Psi_{j}^*) \alpha_N u_j + \dfrac{1}{2} \alpha_N^2 u_j^2 p''_{\lambda_N}(\Psi_{j}^*)(1 + o_{\mathbb{P}}(1)) \right] \\
&=& \alpha_N \left( \dfrac{\partial \ell_N(\Psi^*)}{\partial \Psi_{D^*}} \right)^\top u + \dfrac{N}{2} \alpha_N^2 u^\top \dfrac{1}{N} \dfrac{\partial^2 \ell_N(\Psi^*)}{\partial \Psi_{D^*} \partial \Psi_{D^*}^\top} u (1 + o_{\mathbb{P}}(1)) \\ 
&&- N \dsum_{j=4}^{s_p(D^*)+2} \left[ p'_{\lambda_N}(\Psi_{j}^*) \alpha_N u_j + \dfrac{1}{2} \alpha_N^2 u_j^2 p''_{\lambda_N}(\Psi_{j}^*)(1 + o_{\mathbb{P}}(1)) \right] \\
&=& \alpha_N \left( \dfrac{\partial \ell_N(\Psi^*)}{\partial \Psi_{D^*}} \right)^\top u + \dfrac{N}{2} \alpha_N^2 u^\top \left[ \mathbb{E}\left( \dfrac{1}{N} \dfrac{\partial^2 \ell_N(\Psi^*)}{\partial \Psi_{D^*} \partial \Psi_{D^*}^\top} \right) + o_{\mathbb{P}}(1) \right] u (1 + o_{\mathbb{P}}(1)) \\ 
&&- N \dsum_{j=4}^{s_p(D^*)+2} \left[ p'_{\lambda_N}(\Psi_{j}^*) \alpha_N u_j + \dfrac{1}{2} \alpha_N^2 u_j^2 p''_{\lambda_N}(\Psi_{j}^*)(1 + o_{\mathbb{P}}(1)) \right] \text{ by Lemma \ref{lemme2}} \\
&=& \alpha_N \left( \dfrac{\partial \ell_N(\Psi^*)}{\partial \Psi_{D^*}} \right)^\top u + \dfrac{N}{2} \alpha_N^2 u^\top \mathbb{E}\left( \dfrac{1}{N} \dfrac{\partial^2 \ell_N(\Psi^*)}{\partial \Psi_{D^*} \partial \Psi_{D^*}^\top} \right) u (1 + o_{\mathbb{P}}(1)) + \dfrac{N}{2} \alpha_N^2 u^\top u \ o_{\mathbb{P}}(1) \\ 
&&- N \dsum_{j=4}^{s_p(D^*)+2} \left[ p'_{\lambda_N}(\Psi_{j}^*) \alpha_N u_j + \dfrac{1}{2} \alpha_N^2 u_j^2 p''_{\lambda_N}(\Psi_{j}^*)(1 + o_{\mathbb{P}}(1)) \right] \\
&=& J_1 + J_2 - J_3,
\end{eqnarray*}

with 
$$ \mathbb{E}\left( \dfrac{\partial^2 \ell_N(\Psi^*)}{\partial \Psi_{D^*} \partial \Psi_{D^*}^\top} \right) = 
\begin{pmatrix}
\dfrac{-N}{2 \sigma^{4*}} & \dfrac{-\text{tr}(G_N)}{\sigma^{2*}} & 0 \\
* & -\text{tr}((G_N+G_N^\top)G_N) - \dfrac{1}{\sigma^{2*}}(G_N \chi_N \gamma^*)^\top(G_N \chi_N \gamma^*) & \dfrac{-1}{\sigma^{2*}} \gamma^{*\top} \chi_N^\top G_N^\top \chi_N \\
* & * & \dfrac{-1}{\sigma^{2*}} \chi_N^\top \chi_N
\end{pmatrix},$$
$J_1 = \alpha_N \left( \dfrac{\partial \ell_N(\Psi^*)}{\partial \Psi_{D^*}} \right)^\top u,$
$J_2 = \dfrac{1}{2} \alpha_N^2 u^\top \mathbb{E}\left( \dfrac{\partial^2 \ell_N(\Psi^*)}{\partial \Psi_{D^*} \partial \Psi_{D^*}^\top} \right) u (1 + o_{\mathbb{P}}(1)) + \dfrac{N}{2} \alpha_N^2 u^\top u \ o_{\mathbb{P}}(1)$ and \\
$J_3 = N \dsum_{j=4}^{s_p(D^*)+2} \left[ p'_{\lambda_N}(\Psi_{j}^*) \alpha_N u_j + \dfrac{1}{2} \alpha_N^2 u_j^2 p''_{\lambda_N}(\Psi_{j}^*)(1 + o_{\mathbb{P}}(1)) \right].$ \\

\begin{flalign*}
J_1 &= \alpha_N \left( \dfrac{\partial \ell_N(\Psi^*)}{\partial \Psi_{D^*}} \right)^\top u \\
&= \left\langle \alpha_N \dfrac{\partial \ell_N(\Psi^*)}{\partial \Psi_{D^*}}, u \right\rangle & \\
&\le \bigg|\bigg| \alpha_N \dfrac{\partial \ell_N(\Psi^*)}{\partial \Psi_{D^*}}  \bigg|\bigg| \ ||u|| \text{ by Cauchy–Schwarz inequality} & \\
&= |\alpha_N| \bigg|\bigg| \dfrac{\partial \ell_N(\Psi^*)}{\partial \Psi_{D^*}}  \bigg|\bigg| \ ||u|| & \\
&= O_{\mathbb{P}}(\sqrt{N} \alpha_N) ||u|| \text{ by Lemma \ref{lemme1}} \\
&= O_{\mathbb{P}}(N \alpha_N^2) ||u||.
\end{flalign*}

\begin{flalign*}
J_2 &=  \dfrac{1}{2} \alpha_N^2 u^\top \mathbb{E}\left( \dfrac{\partial^2 \ell_N(\Psi^*)}{\partial \Psi_{D^*} \partial \Psi_{D^*}^\top} \right) u (1 + o_{\mathbb{P}}(1)) + \dfrac{N}{2} \alpha_N^2 u^\top u \ o_{\mathbb{P}}(1) & \\
&=  \dfrac{N}{2} \alpha_N^2 u^\top \dfrac{1}{N} \mathbb{E}\left( \dfrac{\partial^2 \ell_N(\Psi^*)}{\partial \Psi_{D^*} \partial \Psi_{D^*}^\top} \right) u (1 + o_{\mathbb{P}}(1)) + o_{\mathbb{P}}(N \alpha_N^2) ||u||^2 & \\
&= ||u||^2 O_{\mathbb{P}}(N \alpha_N^2 ) \text{ by Assumption I(viii)}. & \\
\end{flalign*}

\begin{flalign*}
J_3 &=  N \dsum_{j=4}^{s_p(D^*)+2} \left[ p'_{\lambda_N}(\Psi_{j}^*) \alpha_N u_j + \dfrac{1}{2} \alpha_N^2 u_j^2 p''_{\lambda_N}(\Psi_{j}^*)(1 + o_{\mathbb{P}}(1)) \right] & \\
&\le N \dsum_{j=4}^{s_p(D^*)+2} \left[ a_N \alpha_N |u_j| + \dfrac{1}{2} \alpha_N^2 u_j^2 b_N (1 + o_{\mathbb{P}}(1)) \right] & \\
&\le N a_N \alpha_N \dsum_{j=1}^{s_p(D^*)+2}  |u_j| +   \dfrac{N}{2} \alpha_N^2 b_N \dsum_{j=1}^{s_p(D^*)+2} u_j^2  (1 + o_{\mathbb{P}}(1)) & \\
&\le  O_{\mathbb{P}}(N \alpha_N^2) ||u|| + o_{\mathbb{P}}(N \alpha_N^2) ||u||^2 .
\end{flalign*}

Thus when $||u|| = C$ is sufficiently large, $J_1$ and $J_3$ are uniformly dominated by $J_2$ which implies (\ref{equation}) 
and Theorem \ref{th1}.
\begin{flushright} $\blacksquare$ \end{flushright}

\section{Proof of Theorem \ref{th2}}

\begin{flalign*}
\dfrac{\partial Q_N(\Psi_{D^*})}{\partial \Psi_{j}} &= \dfrac{\partial \ell_N(\Psi_{D^*})}{\partial \Psi_{j}} - N p'_{\lambda_N}(\Psi_j) \mathds{1}_{j \ge 4}.
\end{flalign*}
By using a Taylor expansion of $\dfrac{\partial \ell_N(\Psi_{D^*})}{\partial \Psi_{j}}$ and $p'_{\lambda_N}(\Psi_j)$ at order 1 we get:
\begin{eqnarray*}
\dfrac{\partial Q_N(\Psi_{D^*})}{\partial \Psi_{j}} &=& \dfrac{\partial \ell_N(\Psi^*)}{\partial \Psi_{j}} + \dsum_{k=1}^{s_p(D^*)+2} \left[ \dfrac{\partial^2 \ell_N(\Psi^*)}{\partial \Psi_k \partial \Psi_j} (\Psi_k - \Psi_{k}^*) \right] + o_{\mathbb{P}}(||\Psi_{D^*} - \Psi^*||) \\
&& - N \left[ p'_{\lambda_N}(\Psi_{j}^*) + p''_{\lambda_N}(\Psi_{j}^*) (\Psi_j - \Psi_{j}^*) + o_{\mathbb{P}}(\Psi_j - \Psi_{j}^*) \right] \mathds{1}_{j \ge 4}.
\end{eqnarray*}
For $\hat{\Psi}_{D^*}$ such that $\dfrac{\partial Q_N(\hat{\Psi}_{D^*})}{\partial \Psi_{j}}=0$ for all $j \in \{ 1, \dots, s_p(D^*)+2 \}$, we have

\begin{eqnarray*}
\dfrac{\partial \ell_N(\Psi^*)}{\partial \Psi_{j}} &=& - \dsum_{k=1}^{s_p(D^*)+2} \left[ \dfrac{\partial^2 \ell_N(\Psi^*)}{\partial \Psi_k \partial \Psi_j} (\hat{\Psi}_k - \Psi_{k}^*) \right] + o_{\mathbb{P}}(||\Psi_{D^*} - \Psi^*||) \\
&& + N \left[ p'_{\lambda_N}(\Psi_{j}^*) + p''_{\lambda_N}(\Psi_{j}^*) (\hat{\Psi}_j - \Psi_{j}^*) + o_{\mathbb{P}}(\hat{\Psi}_j - \Psi_{j}^*)  \right] \mathds{1}_{j \ge 4} \\
&=& - \dsum_{k=1}^{s_p(D^*)+2} \dfrac{\partial^2 \ell_N(\Psi^*)}{\partial \Psi_k \partial \Psi_j} (\hat{\Psi}_k - \Psi_{k}^*) + o_{\mathbb{P}}(||\Psi_{D^*} - \Psi^*||) \\
&& + N p'_{\lambda_N}(\Psi_{j}^*) \mathds{1}_{j \ge 4} + N p''_{\lambda_N}(\Psi_{j}^*) (\hat{\Psi}_j - \Psi_{j}^*) \mathds{1}_{j \ge 4} + o_{\mathbb{P}}(N (\hat{\Psi}_j - \Psi_{j}^*) ) \mathds{1}_{j \ge 4} . \\
\end{eqnarray*}
Since $o_{\mathbb{P}}(||\Psi_{D^*} - \Psi^*||) = o_{\mathbb{P}}(N^{-1/2} + a_N)$ and $o_{\mathbb{P}}(N (\hat{\Psi}_j - \Psi_{j}^*) ) = o_{\mathbb{P}}(\sqrt{N} + N a_N) = o_{\mathbb{P}}(\sqrt{N})$ 
from Theorem \ref{th1} and assumptions of Theorem \ref{th2}, we have 

\begin{eqnarray*}
\dfrac{1}{\sqrt{N}} \dfrac{\partial \ell_N(\Psi^*)}{\partial \Psi_{j}} &=& - \dsum_{k=1}^{s_p(D^*)+2} \dfrac{\partial^2 \ell_N(\Psi^*)}{\partial \Psi_k \partial \Psi_j} \dfrac{1}{N} \sqrt{N} (\hat{\Psi}_k - \Psi_{k}^*) \\
&& + \sqrt{N} p'_{\lambda_N}(\Psi_{j}^*) \mathds{1}_{j \ge 4} + \sqrt{N} p''_{\lambda_N}(\Psi_{j}^*) (\hat{\Psi}_j - \Psi_{j}^*) \mathds{1}_{j \ge 4} + o_{\mathbb{P}}(1 ) . \\
\end{eqnarray*}

Let $A = - \dfrac{1}{N} \dfrac{\partial^2 \ell_N(\Psi^*)}{\partial \Psi_{D^*} \partial \Psi_{D^*}^\top}$, $P_1 = \left(0,0,0,p'_{\lambda_N}(\Psi_{4}^*), \dots, p'_{\lambda_N}(\Psi_{s_p(D^*)+2}^*)\right)^\top$ and \\ $P_2 = \text{diag}\left(0,0,0,p''_{\lambda_N}(\Psi_{4}^*), \dots, p''_{\lambda_N}(\Psi_{s_p(D^*)+2}^*)\right)$. Then

\begin{flalign*}
\dfrac{1}{\sqrt{N}} \dfrac{\partial \ell_N(\Psi^*)}{\partial \Psi_{D^*}} &= \sqrt{N} A (\hat{\Psi}_{D^*} - \Psi^*) + \sqrt{N} P_1 + \sqrt{N} P_2 (\hat{\Psi}_{D^*} - \Psi^*) + o_{\mathbb{P}}(1) & \\
&=  \sqrt{N} \mathbb{E}(A) (\hat{\Psi}_{D^*} - \Psi^*) + \sqrt{N} [A-\mathbb{E}(A)] (\hat{\Psi}_{D^*} - \Psi^*) + \sqrt{N} P_1 + \sqrt{N} P_2 (\hat{\Psi}_{D^*} - \Psi^*) + o_{\mathbb{P}}(1 ) . & \\
\end{flalign*}
From Lemma \ref{lemme2} and Theorem \ref{th1}, we have 
\begin{equation} \label{eqfinal}
\begin{split}
\dfrac{1}{\sqrt{N}} \dfrac{\partial \ell_N(\Psi^*)}{\partial \Psi_{D^*}} &=  \sqrt{N} \mathbb{E}(A) (\hat{\Psi}_{D^*} - \Psi^*) + \sqrt{N} P_1 + \sqrt{N} P_2 (\hat{\Psi}_{D^*} - \Psi^*) + o_{\mathbb{P}}(1) \\
&=  \sqrt{N} \left[(\mathbb{E}(A)+P_2) (\hat{\Psi}_{D^*} - \Psi^*) + P_1 \right] + o_{\mathbb{P}}(1) . \\
\end{split}
\end{equation}

From (\ref{eq1}), (\ref{eq2}) and (\ref{eq3}), $\dfrac{1}{\sqrt{N}} \dfrac{\partial \ell_N(\Psi^*)}{\partial \Psi_{D^*}}$ can also be written as
\begin{flalign*}
\dfrac{1}{\sqrt{N}} \dfrac{\partial \ell_N(\Psi^*)}{\partial \Psi_{D^*}} &= \begin{pmatrix}
    \dfrac{1}{2 \sigma^{4*} \sqrt{N}}[\bm{\varepsilon}_N^\top \bm{\varepsilon}_N - N \sigma^{2*}] \\
    \dfrac{1}{\sqrt{N} \sigma^{2*}} (G_N \chi_N \gamma^*)^\top \bm{\varepsilon}_N + \dfrac{1}{\sqrt{N} \sigma^{2*}} [\bm{\varepsilon}_N^\top G_N \bm{\varepsilon}_N - \sigma^{2*} \text{tr}(G_N)] \\
    \dfrac{1}{\sqrt{N} \sigma^{2*}} \chi_N^\top \bm{\varepsilon}_N
\end{pmatrix} & \\
&= \begin{pmatrix}
    0 \\
    \dfrac{1}{\sqrt{N} \sigma^{2*}} (G_N \chi_N \gamma^*)^\top \bm{\varepsilon}_N \\
    \dfrac{1}{\sqrt{N} \sigma^{2*}} \chi_N^\top \bm{\varepsilon}_N
\end{pmatrix} + \begin{pmatrix}
    \dfrac{1}{2 \sigma^{4*} \sqrt{N}}[\bm{\varepsilon}_N^\top \bm{\varepsilon}_N - N \sigma^{2*}] \\
    \dfrac{1}{\sqrt{N} \sigma^{2*}} [\bm{\varepsilon}_N^\top G_N \bm{\varepsilon}_N - \sigma^{2*} \text{tr}(G_N)] \\
    0
\end{pmatrix} & \\
&= \mathcal{M}_{N,1} + \mathcal{M}_{N,2} . &
\end{flalign*}

Then the central limit theorem for linear-quadratic forms \citep{kelejian2001asymptotic} states that
$$ \dfrac{1}{\sqrt{N}} \dfrac{\partial \ell_N(\Psi^*)}{\partial \Psi_{D^*}} \overset{d}{\underset{N \rightarrow \infty}{\longrightarrow}} \mathcal{N}\left( 0, \underset{N \rightarrow \infty}{\lim} \ \mathbb{V}(\mathcal{M}_{N,1}) + \mathbb{V}(\mathcal{M}_{N,2}) + \mathbb{E}\left(\mathcal{M}_{N,1}\mathcal{M}_{N,2}^\top\right) + \mathbb{E}\left(\mathcal{M}_{N,1}\mathcal{M}_{N,2}^\top\right)^\top \right), $$

where

\begin{flalign*}
\mathbb{V}(\mathcal{M}_{N,1}) &= \begin{pmatrix}
    0 & 0 & 0 \\
    0 & \dfrac{1}{N \sigma^{2*}} (G_N \chi_N \gamma^*)^\top (G_N \chi_N \gamma^*) & \dfrac{1}{N \sigma^{2*}} \gamma^{*\top} \chi_N^\top G_N^\top \chi_N \\
    0 & * & \dfrac{1}{N \sigma^{2*}} \chi_N^\top \chi_N
\end{pmatrix}, &
\end{flalign*}

\begin{flalign*}
\mathbb{V}(\mathcal{M}_{N,2}) &=  
\begin{pmatrix}
\dfrac{\mathbb{E}(\varepsilon_1^4) - \sigma^{4*}}{4 \sigma^{8*}} & \dfrac{\mathbb{E}(\varepsilon_1^4) - \sigma^{4*}}{2N \sigma^{6*}} \text{tr}(G_N) & 0 \\
 * & \dfrac{\mathbb{E}(\varepsilon_1^4)-3\sigma^{4*}}{N \sigma^{4*}}  \dsum_{i=1}^N G_{ii,N}^2 + \dfrac{1}{N} \text{tr}((G_N^\top + G_N)G_N) & 0 \\
 0 & 0 & 0
\end{pmatrix}
&
\end{flalign*}

and

\begin{flalign*}
\mathbb{E}\left(\mathcal{M}_{N,1}\mathcal{M}_{N,2}^\top\right) &=    
\begin{pmatrix}
0 & 0 & 0 \\
\dfrac{\mathbb{E}(\varepsilon_1^3)}{2N \sigma^{6*}} (G_N \chi_N \gamma^*)^\top \mathbf{1}_N & \dfrac{\mathbb{E}(\varepsilon_1^3)}{N \sigma^{4*}} (G_N \chi_N \gamma^*)^\top (G_{11,N}, \dots, G_{NN,N})^\top & 0 \\
\dfrac{\mathbb{E}(\varepsilon_1^3)}{2N \sigma^{6*}} \chi_N^\top \mathbf{1}_N & \dfrac{\mathbb{E}(\varepsilon_1^3)}{\sigma^{4*} N} \chi_N^\top (G_{11,N}, \dots, G_{NN,N})^\top & 0
\end{pmatrix}
& \\
&= \begin{pmatrix}
0 & 0 & 0 \\
\dfrac{\mathbb{E}(\varepsilon_1^3)}{2N \sigma^{6*}} (G_N \chi_N \gamma^*)^\top \mathbf{1}_N & \dfrac{\mathbb{E}(\varepsilon_1^3)}{N \sigma^{4*}} \dsum_{i=1}^N G_{ii,N} G_{i.,N} \chi_N \gamma^* & 0 \\
\dfrac{\mathbb{E}(\varepsilon_1^3)}{2N \sigma^{6*}} \chi_N^\top \mathbf{1}_N & \dfrac{\mathbb{E}(\varepsilon_1^3)}{\sigma^{4*} N} \dsum_{i=1}^N G_{ii,N} \chi_{i.,N}^\top & 0
\end{pmatrix} &
\end{flalign*}
where $G_{i.,N}$ and $\chi_{i.,N}$ denote the $i^\text{th}$ row of $G_N$ and $\chi_N$ respectively.

We conclude by (\ref{eqfinal}) and Slutsky's theorem:
\begin{align*}
  \sqrt{N} [(\mathbb{E}(A)+P_2) (\hat{\Psi}_{D^*} - \Psi^*) + P_1] \overset{d}{\underset{N \rightarrow \infty}{\longrightarrow}} \mathcal{N}\left(0, \underset{N \rightarrow \infty}{\lim} \ \right. &\mathbb{V}(\mathcal{M}_{N,1}) + \mathbb{V}(\mathcal{M}_{N,2})  \\ &\left. + \mathbb{E}\left(\mathcal{M}_{N,1}\mathcal{M}_{N,2}^\top\right) + \mathbb{E}\left(\mathcal{M}_{N,1}\mathcal{M}_{N,2}^\top\right)^\top \right) .   
\end{align*}
\begin{flushright} $\blacksquare$ \end{flushright}

\section{Results of the simulation study for $k=8$} \label{sec:k8}

\begin{figure}
\centering
\includegraphics[width=\textwidth]{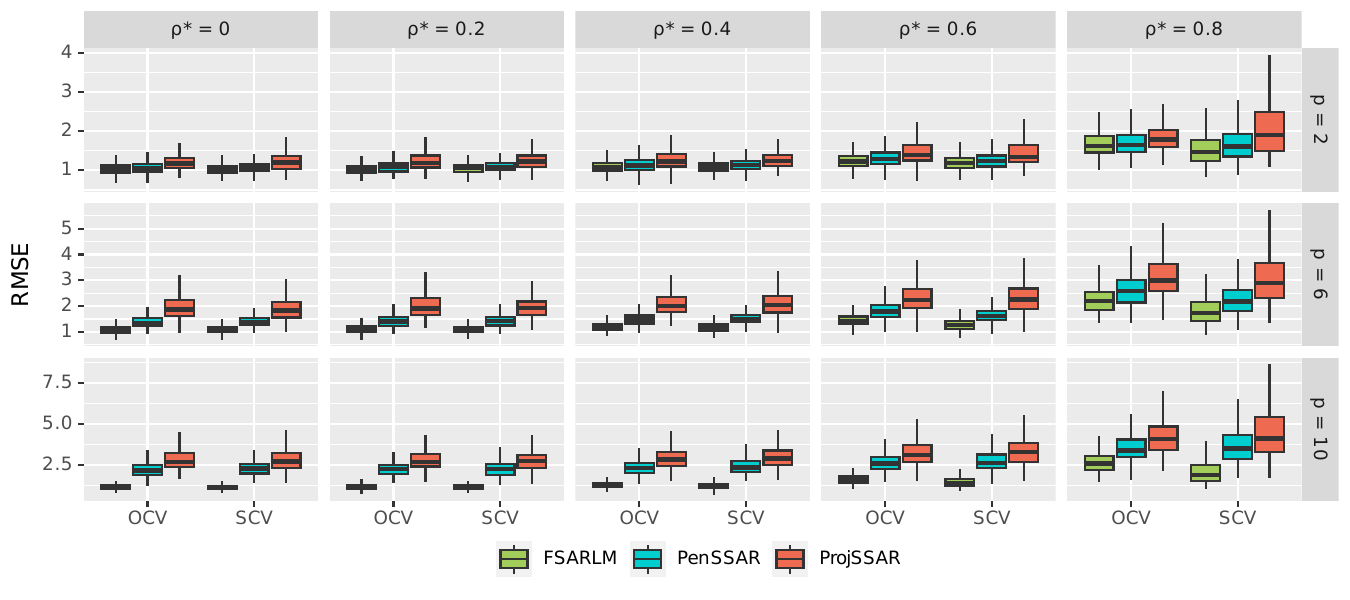}
\caption{RMSE on the test set with the FSARLM, the PenSSAR and the ProjSSAR for Model 1 with $k=8$ using ordinary (OCV) and spatial (SCV) cross-validation}
\label{fig:rmse8_1}
\end{figure}

\begin{figure}
\centering
\includegraphics[width=\textwidth]{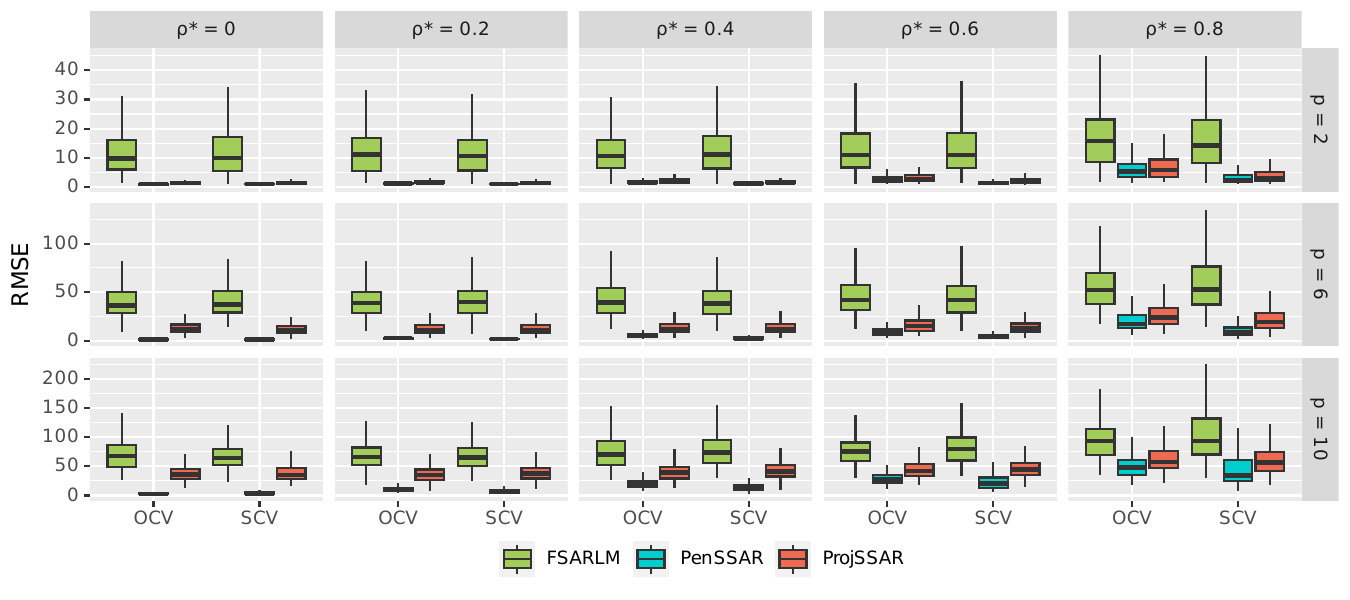}
\caption{RMSE on the test set with the FSARLM, the PenSSAR and the ProjSSAR for Model 2 with $k=8$ using ordinary (OCV) and spatial (SCV) cross-validation}
\label{fig:rmse8_2}
\end{figure}

\begin{figure}
\centering
\includegraphics[width=\textwidth]{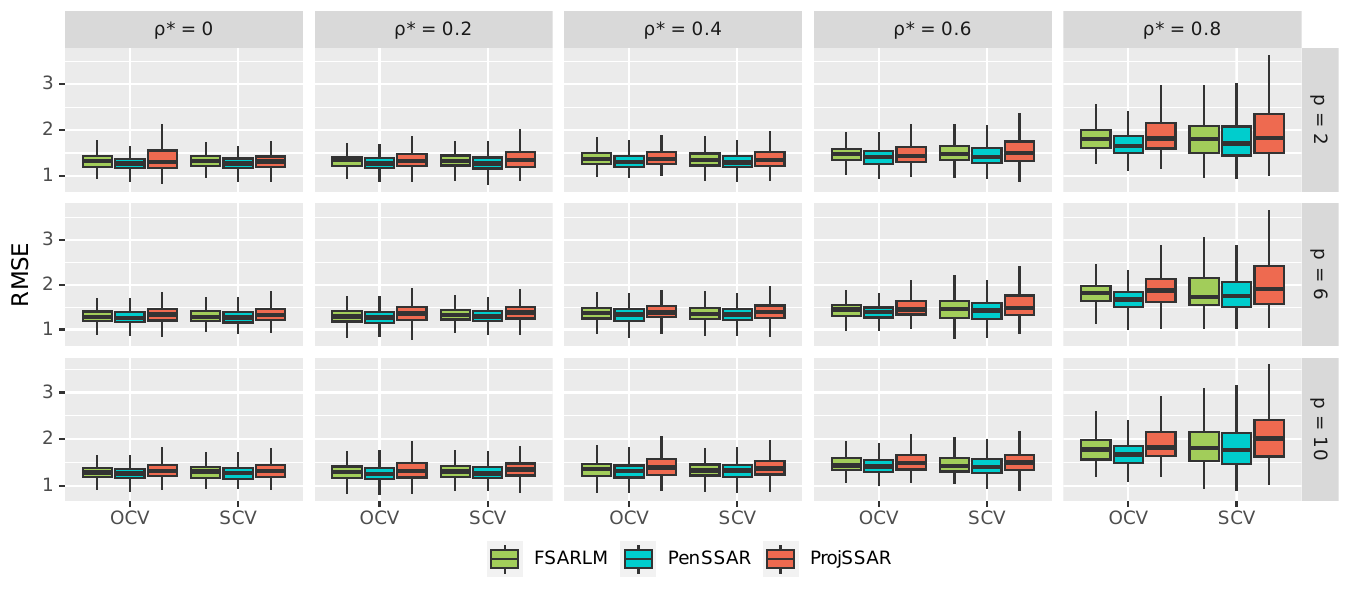}
\caption{RMSE on the test set with the FSARLM, the PenSSAR and the ProjSSAR for Model 3 with $k=8$ using ordinary (OCV) and spatial (SCV) cross-validation}
\label{fig:rmse8_3}
\end{figure}

\begin{figure}
\centering
\includegraphics[width=\textwidth]{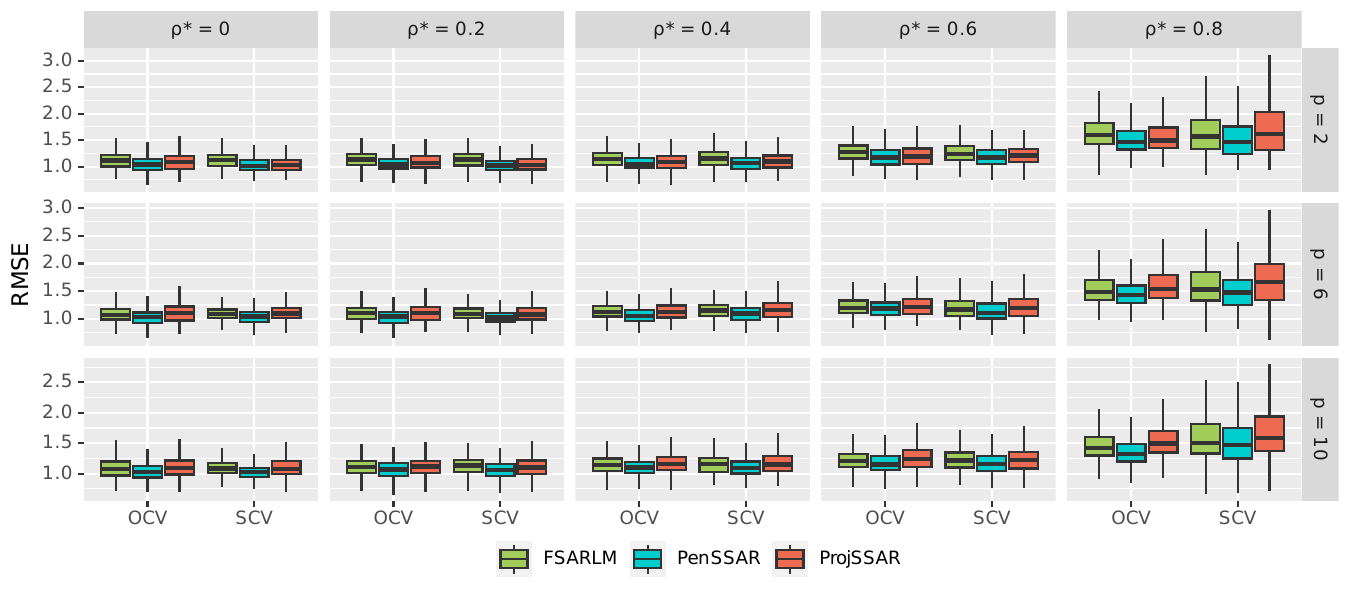}
\caption{RMSE on the test set with the FSARLM, the PenSSAR and the ProjSSAR for Model 4 with $k=8$ using ordinary (OCV) and spatial (SCV) cross-validation}
\label{fig:rmse8_4}
\end{figure}

\begin{figure}
\centering
\includegraphics[width=\textwidth]{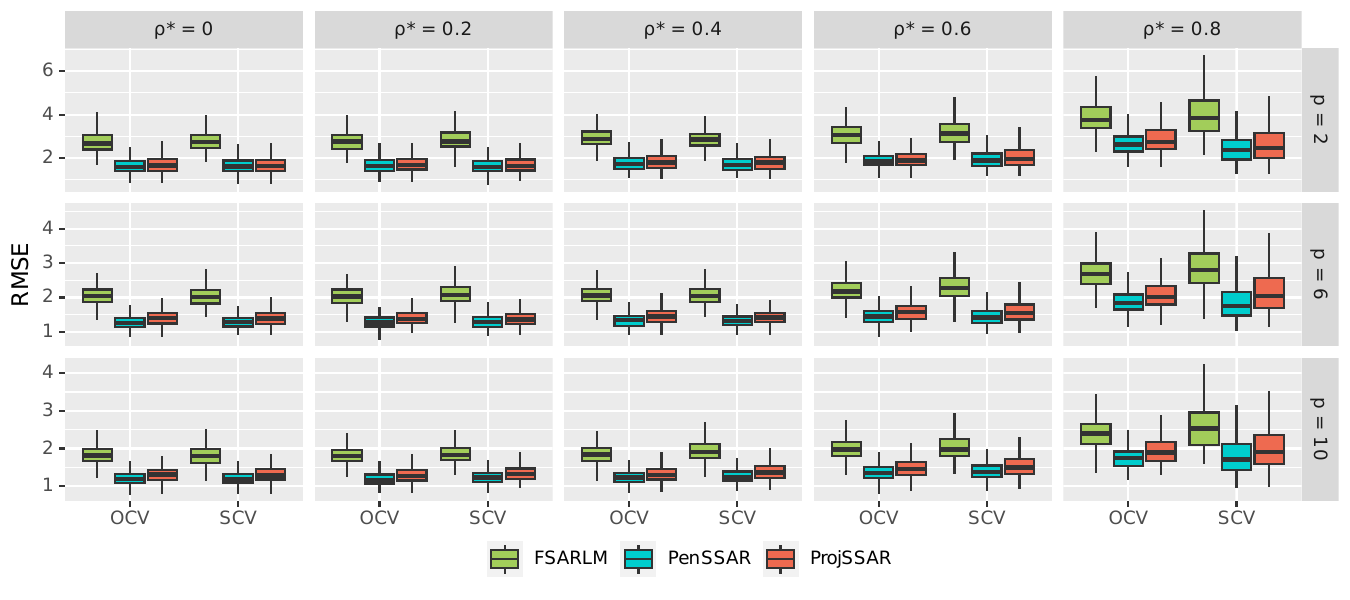}
\caption{RMSE on the test set with the FSARLM, the PenSSAR and the ProjSSAR for Model 5 with $k=8$ using ordinary (OCV) and spatial (SCV) cross-validation}
\label{fig:rmse8_5}
\end{figure}

\end{document}